\documentclass[twocolumn]{aastex631}

\usepackage{multirow}
\usepackage{comment}
\usepackage{amsmath}
\usepackage{diagbox}

\usepackage{CJKutf8}            

\begin{document}

\title{A Local Dwarf Galaxy Search Using Machine Learning}

\correspondingauthor{Huanian Zhang, Guangping Ye}
\email{huanian@hust.edu.cn, guangping@hust.edu.cn}

\author[0000-0002-0123-9246]{Huanian Zhang \begin{CJK*}{UTF8}{gbsn} (张华年) \end{CJK*}}
\affiliation{Department of Astronomy, Huazhong University of Science and Technology, Wuhan, Hubei 430074, China}
\affiliation{Steward Observatory, University of Arizona, Tucson, AZ 85719, USA}

\author{Guangping Ye \begin{CJK*}{UTF8}{gbsn} (叶广平) \end{CJK*}}
\affiliation{Department of Astronomy, Huazhong University of Science and Technology, Wuhan, Hubei 430074, China}

\author{Rongyu Wu \begin{CJK*}{UTF8}{gbsn} (吴嵘玉) \end{CJK*}}
\affiliation{Department of Astronomy, Huazhong University of Science and Technology, Wuhan, Hubei 430074, China}

\author[0000-0002-5177-727X]{Dennis Zaritsky}
\affiliation{Steward Observatory, University of Arizona, Tucson, AZ 85719, USA}





\begin{abstract}

We present a machine learning search for local, low-mass  galaxies ($z < 0.02$ and $10^6 M_\odot < M_* < 10^9 M_\odot$) using the combined photometric data from the DESI Imaging Legacy Surveys and the WISE survey. We introduce the spectrally confirmed training sample, discuss evaluation metrics, investigate the features, compare different machine learning algorithms, and find that a 7-class neural network classification model is highly effective in separating the signal (local, low-mass  galaxies) from various contaminants, reaching a precision of 95\% and a recall of 76\%. The principal contaminants are nearby sub-$L^*$ galaxies at $0.02 < z < 0.05$ and nearby massive galaxies at $0.05 < z < 0.2$. We find that the features encoding surface brightness information are essential to achieving a correct classification.
Our final catalog, which we make available, consists of 112,859 local, low-mass  galaxy candidates, where 36,408 have high probability ($p_{\rm signal} > 0.95$), covering the entire Legacy Surveys DR9 footprint.
Using DESI-EDR public spectra and data from the SAGA and ELVES surveys, we 
find that our model has a precision of $\sim 100\%$, $96\%$, and $97\%$, respectively, 
and a recall of \textbf{$\sim 51\%$,} $68\%$ and $53\%$, respectively.  The results of those independent spectral verification demonstrate the effectiveness and efficiency of our machine learning classification model. 
\end{abstract}



\section{Introduction} \label{sec:intro}

Low-mass ($M_* < 10^9$ M$_\odot$) galaxies, often referred to as dwarf galaxies \citep[e.g.][]{Mateo1998, Tolstoy2009}, provide an excellent laboratory for the study of a wide range of scientific phenomena and are the most numerous type of galaxies in the Universe \citep{Schechter1976}. The halos of these galaxies are generally dark matter (DM) dominated \citep[e.g.][]{Simon2019}, providing an important window into the physics of dark matter by allowing us to trace the dark matter distribution on small scales \citep[e.g.][]{Karukes2017}. Furthermore, their shallow gravitational potentials make them sensitive to ``feedback" \citep[e.g.][]{Reines2013, Wasleske2024} and a range of environmental effects \citep[e.g.][]{Deason2014, Paudel2023, Christensen2024}. Their relatively simple formation histories make it easier to identify certain aspects of galaxy formation, such as the effects of reionization \citep{McQuinn2024}. Lastly, local, late-type, ultra low-mass galaxies might be analogs of galaxies at early Universe that can be studied in detail.

Although dwarf galaxies are the most abundant type of galaxy, they can be difficult to find and characterize. They contribute a minority (a few percent) of the local luminosity and stellar mass density \citep[e.g.][]{Bernstein1995, Driver1999, Hayward2005, Martin2019} and tend to have low surface brightness \citep{Bennet2017, Carlsten2021, Jackson2021}. The limiting step in their study is generally the spectroscopic follow-up needed to measure a distance and determine the galaxies' physical parameters. Therefore, many searches, even those with spectroscopic follow-up, have focused on environments where there is a positive bias that the identified candidates are indeed dwarf satellites of massive hosts. Examples include searches for satellite galaxies around massive galaxies \citep[eg.][]{Zaritsky1993, Zaritsky1997, Spencer2014, Kondapally2018, Smercina2018, Tanaka2018}.  The  current, state-of-the-art searches include those for satellites of Milky Way analogs, called ``Satellites Around Galactic Analogs" \citep[SAGA,][]{SAGA2017} and for satellites of massive hosts in the Local Volume, called ``Exploration of Local VolumE Satellites" \citep[ELVES,][]{ELVES2022}. Searches for dwarf galaxies in even denser environments, such as galaxy groups and clusters, include the Next Generation Virgo Cluster Survey \citep[NGVS,][]{Ferrarese2012}, the Next Generation Fornax Survey \citep[NGFS,][]{Eigenthaler2018}, and the SAMI-Fornax dwarf survey \citep{Scott2020}.
However, such focused searches do not span the full range of environment.

The traditional method to search for low-mass galaxies from imaging surveys utilizes the measured colors and photometric half light radii \citep{Greco2018}. Therefore, the traditional method is not optimized  to use all of the available data. New, deeper, public imaging surveys are creating opportunities for systematic searches for local, low-mass galaxies across most of the sky. For example, the Dark Energy Spectroscopic Instrument (DESI) Legacy Imaging Surveys \citep[hereafter referred to as the Legacy Survey or simply LS;][]{Dey2019} deliver deep images in three broad bands for a large area of sky. 
In concert with these new data, the ability of machine learning-based methods to make full use of the photometric measurements, construct correlations in the high-dimensional space, and handle far more data than what can be done manually, allows us to exploit these new opportunities. 

There are various examples of machine learning algorithm-based searches for low surface brightness (LSB) galaxies in the Local Universe \citep[eg.][]{SMUDGes2019, SMUDGes2021, SMUDGes2022, SMUDGes2023, Tanoglidis2021a, Tanoglidis2021b, Muller2021} using the Legacy Survey or the Dark Energy Survey DR1 \citep[DES,][]{DES2016, DES2018}, or for nearby extremely metal-poor galaxies with typical stellar mass of $\sim 10^8$ M$_\odot$ \citep{Cheng2025}. 
We expand on such studies by utilizing a sample of spectrally confirmed dwarf galaxies with stellar mass estimates as the training sample with which to construct a machine learning model to search for local, low-mass galaxies. 

We aim to accurately identify as many local ($z < 0.02$) low-mass ($M_* < 10^9$ M$_\odot$) galaxies as possible via machine learning algorithms using both the Legacy Survey \citep{Dey2019} and the {\it Wide-field Infrared Survey Explorer} (WISE) all-sky survey \citep{Wright2010,Mainzer2011}. The structure of the paper is organized as follows. In Sec. \ref{sec:data} we introduce the catalog data we will use in this study and our training sample. In Sec. \ref{sec:ml} we briefly discuss the various aspects of the machine learning algorithms. In Sec. \ref{sec:candidates} we discuss the procedure we use to obtain the local, low-mass galaxy candidates, present the final catalog, and evaluate the results using available redshift measurements. In Sec. \ref{sec:discussion} we discuss the effect of having an imbalanced data set on the machine learning model, potential future improvements in the model performance, and the impact of future imaging surveys. Finally we summarize our results. The magnitudes used in this paper are based on the AB system \citep{OKE1, OKE2} after applying Galactic extinction corrections \citep{Schlegel1998, Fitzpatrick1999, Schlafly2011}. We adopt a $\Lambda$CDM cosmology with $\Omega_\Lambda = 0.7$, $\Omega_m = 0.3$, and $H_0 = 70 \text{ km s}^{-1} \text{Mpc}^{-1}$ \citep[cf.][]{Riess2018,Planck2018}.

\section{Data} \label{sec:data}
\subsection{Legacy Survey Data} \label{sec:ls}

\begin{figure}[ht]
\begin{center}
\includegraphics[width = 0.48 \textwidth]{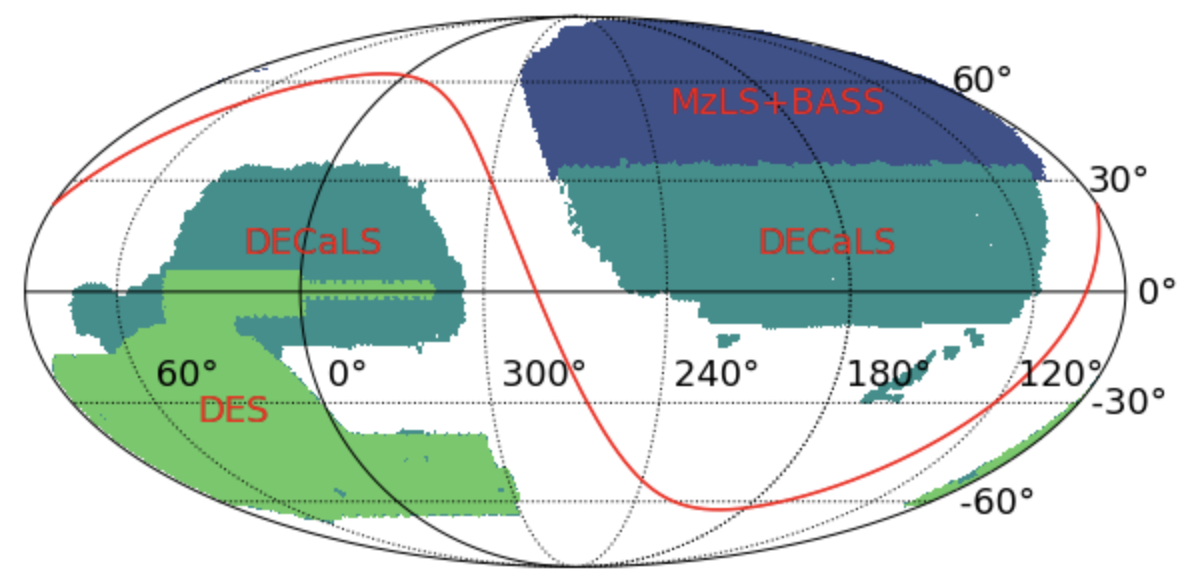}
\end{center}
\caption{ Footprint of the sky covered the Legacy Survey DR9 \citep{Dey2019}. It incorporates the DES in the south. The Galactic plane is traced by the red curve. }
\label{fig:footprint}
\end{figure}

We use the photometric data from the Data Release 9 (DR9) of the Legacy Survey, which includes observations obtained by the DECam at the CTIO 4m (DECaLS), an upgraded MOSAIC camera at the KPNO 4 m telescope (MzLS, Mayall z-band Legacy Survey), and the 90Prime camera \citep{Williams2004} at the Steward Observatory 2.3 m telescope (BASS, Beijing-Arizona Sky Survey). Because the depth of Sloan Digital Sky Survey \citep[SDSS,][]{York2000,Abazajian2009,Aihara2011} is insufficient to provide reliable DESI targets, the Legacy Survey \citep{Dey2019} was initiated to provide targets for the DESI spectral survey \citep{DESI1, DESI2} drawn from deep, three-band ($g=24.7$, $r=23.9$, and $z=23.0$ AB mag, 5$\sigma$ point-source limits) images, roughly two magnitudes deeper than the SDSS \citep{York2000} data. The survey covers about 14,000 deg$^2$ of sky visible from the northern hemisphere between declinations approximately bounded by $-$18$^\circ$ and +84$^\circ$. The footprint of DR9 (Figure \ref{fig:footprint}) also includes an additional 6,000 deg$^2$ extending down to $-$68$^\circ$ imaged at the CTIO by the Dark Energy Survey \citep[DES,][]{DES2016}. 

In addition to the regular photometric measurements, the Legacy Survey catalog also provides fluxes within different apertures, referred to as `apfluxes', which are demonstrated to be highly effective in distinguishing high-redshift quasars from extended objects like galaxies \citep{Ye2024}. We find apfluxes to be highly effective in our modeling because they provide constraints both on the surface brightness profile and color gradients.  The radii of the apertures provided are [0.5$^{\prime\prime}$, 0.75$^{\prime\prime}$, 1.0$^{\prime\prime}$, 1.5$^{\prime\prime}$, 2.0$^{\prime\prime}$, 3.5$^{\prime\prime}$, 5.0$^{\prime\prime}$, 7.0$^{\prime\prime}$] for the {\it g, r, z} bands and [3$^{\prime\prime}$, 5$^{\prime\prime}$, 7$^{\prime\prime}$, 9$^{\prime\prime}$, 11$^{\prime\prime}$] for the $W1, W2$ bands (see \S\ref{sec: WISE} for discussion of the WISE data). We construct ratios of apflux values and use those as machine learning features (see Sec. \ref{sec:Features}). 

\subsection{The Wide-field Infrared Survey Explorer Data} \label{sec: WISE}

The WISE survey provides infrared photometry of four bands at central wavelength of 3.4, 4.6, 12, and 22 $\mu$m ($W1, W2, W3, W4$) over the entire sky \citep{Wright2010, Mainzer2011}. The $W1, W2, W3$ and $W4$ fluxes included in the Legacy Survey catalog are based on forced photometry of the unWISE stacked images (including all imaging through year 7 of NEOWISE-Reactivation) at the locations of the Legacy Surveys optical sources. We use only the $W1$ (3.4 $\mu$m) and $W2$ (4.6 $\mu$m) photometry, which have limiting magnitudes of 19.8 and 19.0 mag in AB magnitudes (17.1 and 15.7 in Vega magnitudes), respectively. The conversion between AB and Vega magnitude for $W1$ and $W2$ is $W1_{\rm AB}$ = $W1_{\rm Vega}$ + 2.699 and $W2_{\rm AB}$ = $W2_{\rm Vega}$ + 3.339 \citep{Jarrett2011}, respectively. Although the stacked WISE images are relatively shallow, the $W1$ and $W2$ photometric measurements are essential to construct a good machine learning model.

\subsection{Training Sample}
\label{sec:TrainingSample}

We use supervised machine learning algorithms to segregate the ``signal" (our target local, dwarf galaxies) from various contaminants. Constructing a good training sample is essential to any successful classification model. Bias in the training sample will propagate to the new dataset, resulting in highly unreliable predictions. To construct a representative training sample spanning the relevant ranges of stellar mass (M$_*$) and redshift ($z$), we collect spectrally confirmed signal (galaxies with $z < 0.02$ and $10^6 < M_* < 10^9$ M$_\odot$) and examples of all possible contaminants that overlap in color/magnitude space or have similar surface brightness profiles. We mainly extract the training sample from the catalog presented by \cite{Chang2015}, which provides stellar masses and star formation rate estimates for over 1 million WISE + SDSS spectroscopic galaxies, and the recent study of local galaxies within 50 Mpc \citep[hereafter called 50Mpc,][]{Ohlson2024}. 
We use these sources to classify objects as signal or contaminants and obtain the photometric and morphological data from both the Legacy Survey ({\it g, r, z} bands) and WISE ($W1, W2$ bands).

\begin{figure}[ht]
\begin{center}
\includegraphics[width = 0.48 \textwidth]{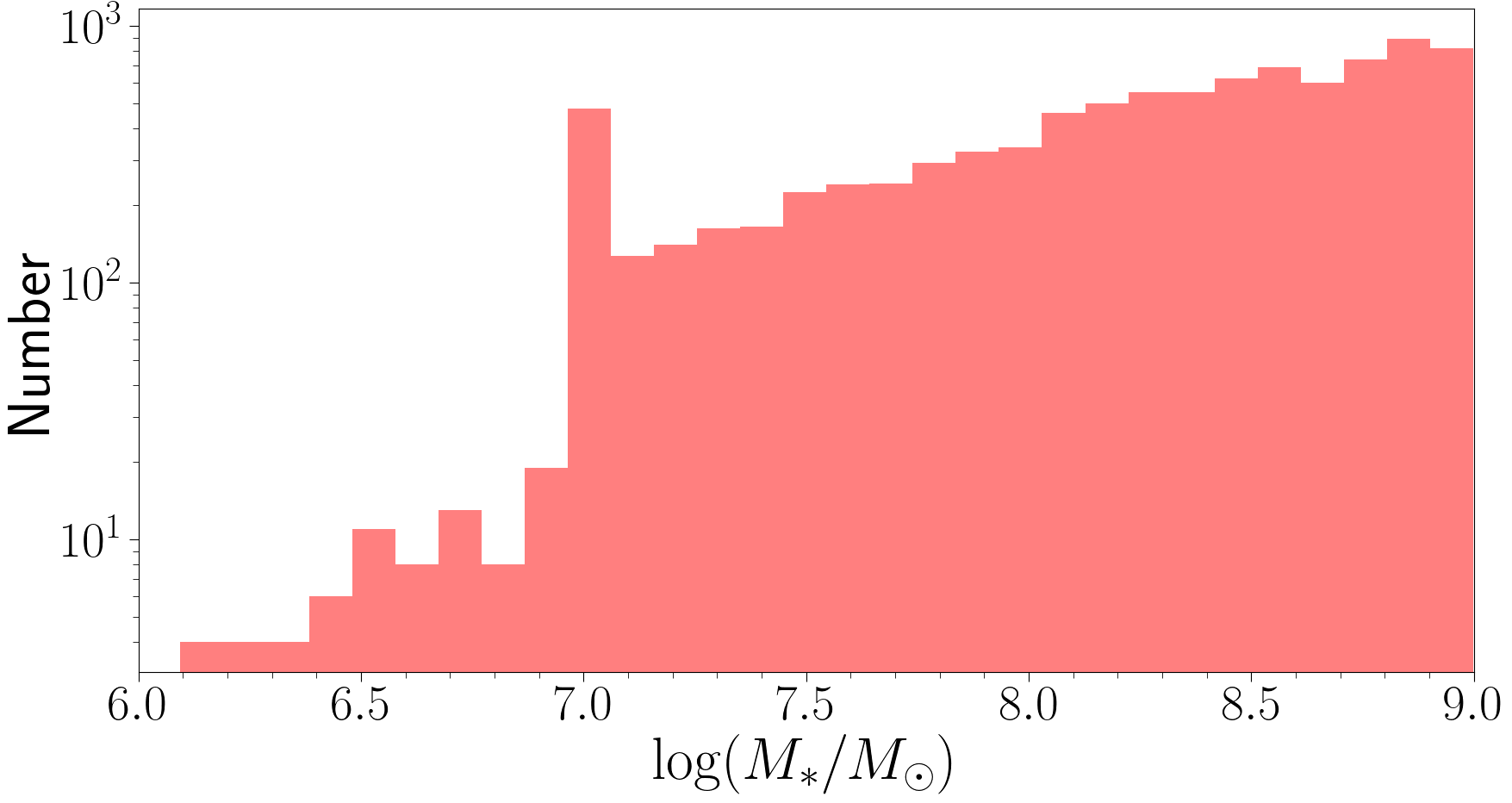}
\end{center}
\caption{The stellar mass distribution of the local low-mass galaxies in the training sample. The reason why there is a spike at $M_* = 10^7$ M$_\odot$ is that $10^7$ M$_\odot$ is the lower stellar mass limit in \cite{Chang2015}. But it does not affect our final results.}
\label{fig:massDis}
\end{figure}

We present the distribution of masses for our ``signal" in Figure \ref{fig:massDis}. There is a sharp drop in number at stellar mass $< 10^7$ M$_\odot$ that is due to survey sensitivity (both in magnitude and surface brightness).
The stellar mass measurements from \cite{Chang2015} are estimated using stellar spectral energy distribution (SED) fitting while those in the 50Mpc catalog are based on the $g–i$ versus $M/L_i$ relation from \cite{Taylor2011}.
We divide the contaminants into a few categories (see Table \ref{tab:training}): (1) nearby sub-$L^*$ galaxies (NLS) with stellar mass of $10^{9.0} < M_* < 10^{9.75}$ at $0.02 < z < 0.05$ are mainly from \cite{Chang2015} 
; (2) local massive galaxies (LM) with $M_* > 10^{9.5}$ M$_\odot$ (or if $M_*$ is unavailable M$_r < -18.8$ mag) and $z < 0.02$ \citep{Chang2015,Sanchez2022, SGA2023, Ohlson2024}; 
(3) nearby massive galaxies (NM)
that are a bit farther $0.05 < z < 0.2$ and but also more massive than the signal, $10^{9.25} < M_* < 10^{10}$ M$_\odot$ \citep{Mendel2014, Geller2014, Drinkwater2018, DuartePuertas2022};(4) higher redshift galaxies (HZ) with $z > 0.2$ and $M_* > 10^{10}$ M$_\odot$ \citep{Geller2014, Drinkwater2018, DuartePuertas2022}; (5) nearby spiral galaxies (NS) at $z > 0.03$ \citep{Kautsch2006, Willett2013, Kourkchi2020}; and (6) ``Artifact Features" (AFs), which include stellar halos, diffraction spikes, Galactic cirrus, tidal ejecta connected to high-surface-brightness host galaxies, knots and star-forming regions in the arms of large spiral galaxies \citep[see][for details]{Tanoglidis2021a}.

\begin{table*}[t]
\centering
\begin{tabular}{|c|c|}
\hline
Category & Number 	  
\\ \hline

Signal (S, $z < 0.02$ and $10^6 < M_* < 10^9$ M$_\odot$) 
& 9,256 
\\ \hline
Nearby Sub-$L^*$ (NSL, $0.02 < z < 0.05$ and $10^{9} < M_* < 10^{9.75}$ M$_\odot$) 
& 29,093
\\ \hline
Local Massive (LM, $z < 0.02$ and $M_* > 10^{9.5}$ M$_\odot$)  
& 3,299  
\\ \hline	
Nearby Massive (NM, $0.05 < z < 0.2$ and $10^{9.25} < M_* < 10^{10}$ M$_\odot$) 
& 65,068  
\\ \hline
High-z Massive (HM, $z > 0.2$ and  $M_* > 10^{10}$ M$_\odot$)    
& 150,398
\\ \hline
Nearby Spiral (NS, $z > 0.03$)     
&  21,294
\\ \hline
Artifact Features (AF)    
&  7,918
\\ \hline											
\end{tabular}
\caption{\label{tab:training}The number of spectrally confirmed signal of local low-mass galaxies as well as the various categories of contaminants in the training sample.}
\end{table*}

\begin{figure}[ht]
\begin{center}
\includegraphics[width = 0.48 \textwidth]{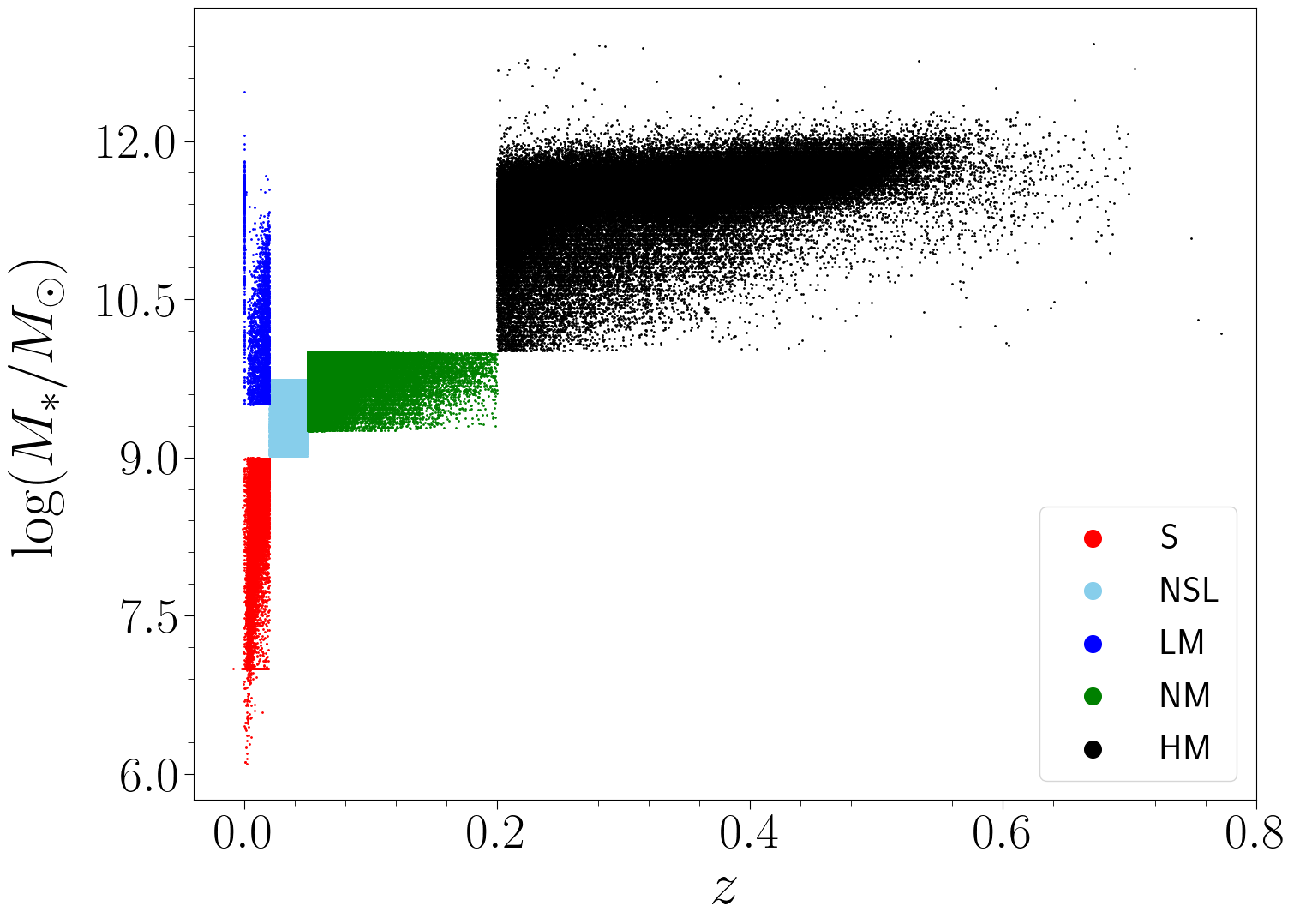}
\includegraphics[width = 0.48 \textwidth]{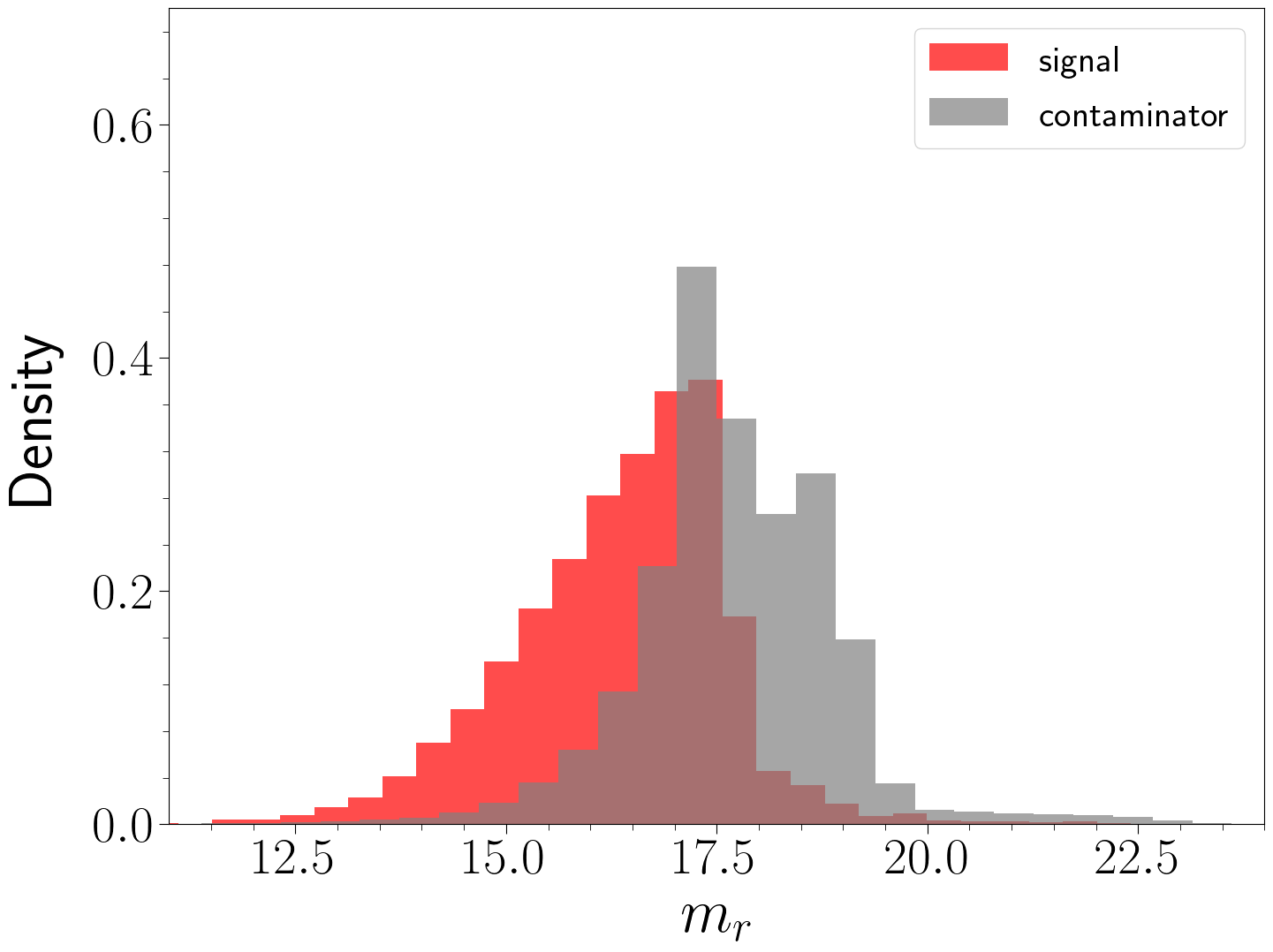}
\end{center}
\caption{The stellar mass vs. redshift of the contaminants in the training sample (top) and the $r$-band magnitude distribution for the signal and all contaminants (bottom).} 
\label{fig:mass_z}
\end{figure}

Both observations \citep[e.g.,][]{Eigenthaler2018, Carlsten2021} and simulations \citep[e.g.,][]{Kaufmann2007, Klein2024} find that the half light radii ($r_{1/2}$) or effective radii ($r_{e}$) of dwarf galaxies vary from several hundred pc to several kpc, with a median value of $(0.67-0.7) \pm 0.02$ kpc \citep{Eigenthaler2018}. This physical size corresponds roughly to $2-3$ arcseconds at the typical distance of our signal in the training sample. Therefore, to obtain a training sample that is as representative as possible, we set a lower limit on the angular half light radius of 1 arcsecond for the entire training sample, which removes $\sim$ 1\% of the signal. This criterion also removes a large fraction of unreliable data in LS DR9 database that we will discuss in detail in \S\ref{sec:candidates}. We present the number of sources in each category in the training sample in Table \ref{tab:training} and the redshift vs. stellar mass, as well as, the $r$ band magnitude distributions for signal and contaminants in Figure \ref{fig:mass_z}. We have no mass estimates for the NS sample, but they might reside in the lower right region in the $M_*-z$ phase space. The strict boundary cut along either redshift or stellar mass will introduce confusion between classes, which will be further discussed in \S\ref{sec:improvement}, but our aim is not to correctly classify subtypes of contaminants. Finally, we note that the number of objects in each class in the training sample is not equal, which means that we are dealing with the imbalanced classification. This issue will be further discussed in \S\ref{sec:imbalance}. Ultimately, however, the test of this method is simply in the accuracy with which we label something as signal and how much signal we fail to identify.

\section{Machine Learning} \label{sec:ml}


In the past few decades, many large surveys have already collected data that is beyond our ability to interactively analyze, and future imaging or spectra surveys will further exacerbate this situation. An approach using automatic methods, like machine learning, is necessary, efficient, and easily reproducible. Here we briefly introduce the various aspects of the machine learning technique.

\subsection{Metrics} \label{sec:Metrics}

Ultimately, any method must prove its worth quantitatively.
We now introduce metrics that we will utilize to evaluate various machine learning algorithms. There is no perfect metric, each metric has its own strengths and weaknesses. 
For our construction of metrics we tabulate the incidence rates of the four possible outcomes of each classification exercise: correctly identifying a detection as signal or a true positive (TP), correctly identifying it as contamination or a true negative (TN),  incorrectly identifying it as signal or a false positive (FP), and incorrectly identifying it as contamination or a false negative (FN). Mathematical combinations of these four measurements provide evaluation metrics, each with a different emphasis:

\medskip
\noindent
{\sl Precision} measures the fraction of the population identified as signal that is actually signal, or
    
\begin{equation}
	\label{eqn:Precision}
	Precision = \frac{TP}{TP + FP}. 
\end{equation}	

\medskip\noindent
{\sl Recall} measures the fraction of the actual signal that is correctly recovered, or

\begin{equation}
	\label{eqn:recall}
	Recall = \frac{TP}{TP + FN} 
\end{equation}

\medskip\noindent
{\sl Accuracy} measures the overall correctness of the classification, regardless of whether the detection is signal or contaminant, or

\begin{equation}
	\label{eqn:accu}
	Accuracy = \frac{TP + TN}{TP + TN + FP + FN} 
\end{equation}	

\medskip\noindent
The {\sl f1 score} is the harmonic mean of precision and recall. This metric aims to measure the overall quality of the classification by weighing both precision and recall. On datasets where the classes are roughly balanced, the f1 score performs well. 

\medskip\noindent
The {\sl Adjusted f-score} (AGF) 
is an improvement upon the f1 score for highly imbalanced datasets \citep{Akosa2017}. AGF considers all of the elements in the confusion matrix (TP, FP, TN, FN), making it a more equitable evaluation metric for the classification of the minority class. Additional discussion of f1 and AGF can be found in \cite{Ye2024}.

Which specific metric one chooses to emphasize depends on the ultimate goal of the study. Our focus is to provide a sample of dwarf galaxy candidates that are most likely to survive spectroscopic confirmation and therefore we emphasize precision. Other investigators may wish to emphasize completeness, and so would place a greater weight on recall.
In principle, precision, recall, f1, and AGF are less affected by an imbalanced data sample than accuracy. We will not use accuracy in this study. Meanwhile, it is almost impossible to tune the desired metrics for all classes for a multi-class classification task. We choose to focus on the metrics of precision, recall, f1 and AGF for the signal (local dwarf galaxies).
At the same time, to not focus solely on the performance of our model on signal, we also provide weighted metrics for the other six classes, which can showcase that our model has a comprehensive performance across these classes of sources.

\subsection{Features} 
\label{sec:Features}

The machine learning algorithm performs using a set of input measurements or ``features". We provide it with dereddened magnitudes which are estimated from a best model fitting and aperture fluxes in the {\it g, r, z}, $W1$ and $W2$ bands. The aperture radii are [0.5$^{\prime\prime}$, 0.75$^{\prime\prime}$, 1.0$^{\prime\prime}$, 1.5$^{\prime\prime}$, 2.0$^{\prime\prime}$, 3.5$^{\prime\prime}$, 5.0$^{\prime\prime}$, 7.0$^{\prime\prime}$] for the {\it g, r, z} bands, and are [3$^{\prime\prime}$, 5$^{\prime\prime}$, 7$^{\prime\prime}$, 9$^{\prime\prime}$, 11$^{\prime\prime}$]  for the $W1$ and $W2$ bands. 
The features used in the machine learning algorithm are the magnitudes, colors  which are constructed from the subtraction of two magnitudes, and flux ratios (apfluxes). 
The apflux ratios are calculated 
as follows: 1) for apfluxes in the same filter, we calculate the ratio of aperture fluxes within apertures with consecutive radii; 2) for apfluxes among different filters but entirely within either the optical bands ({\it g, r, z}) or within the infrared bands ($W1$, $W2$), 
we calculate apflux ratios both among apertures for the same filter passband and across passbands. Table \ref{tab:features} provides a detailed overview of the different features
utilized in our model training.
In total, we use 5 magnitudes, 10 colors, and 76 apflux ratios between combinations of filters and apertures. 

\begin{table*}[t]
\centering
\begin{tabular}{ccl|c}
\hline
Type & \multicolumn{2}{c|}{Features} & Number \\ \hline
\multicolumn{1}{c|}{Magnitudes}                                 
& \multicolumn{2}{c|}{g,r,z,W1 and W2}                            
& 5       \\ \hline
\multicolumn{1}{c|}{Colors}                                    
& \multicolumn{2}{c|}{g-\{r, z, W1, W2\}, r-\{z, W1, W2\}, z-\{W1, W2\} and W1-W2} 
& 10      \\ \hline
\multicolumn{1}{c|}{Apflux ratios (same filters, diff. radii)}    
& \multicolumn{2}{c|}{\begin{tabular}[c]{@{}c@{}}apflux\_\{g,r,z\}\_\{i\}/apflux\_\{g,r,z\}\_\{i+1\} \\ and apflux\_\{W1,W2\}\_\{j\}/apflux\_\{W1,W2\}\_\{j+1\}\end{tabular}} 
& 29      \\ \hline
\multicolumn{1}{c|}{Apflux ratios (diff. filters, same radii)}    
& \multicolumn{2}{c|}{\begin{tabular}[c]{@{}c@{}}apflux\_g\_i/apflux\_\{r,z\}\_i, apflux\_r\_i/apflux\_z\_i \\ and apflux\_W1\_j/apflux\_W2\_j\end{tabular}}          
& 29      \\ \hline
\multicolumn{1}{c|}{Apflux ratios (diff. filters, similar radii)} & \multicolumn{2}{c|}{\begin{tabular}[c]{@{}c@{}}apflux\_\{g,r,z\}\_i/apflux\_W1\_j \\ and apflux\_\{g,r,z\}\_i/apflux\_W2\_j\end{tabular}}                                   
& 18      \\ \hline
\end{tabular}\caption{\label{tab:features} The features in our classification model, including magnitude, color, and apflux ratio features. The index i represents the aperture size sequence for the $g$, $r$, and $z$ filters (i=1 to 8), while the index j represents the aperture size sequence for the W1 and W2 filters (j=1 to 5). The last three aperture sizes of the g, r, and z filters are very close to the first three aperture sizes of the W1 and W2 filters. }
\end{table*}




\begin{figure*}[ht]
\begin{center}
\includegraphics[width = 0.8 \textwidth]{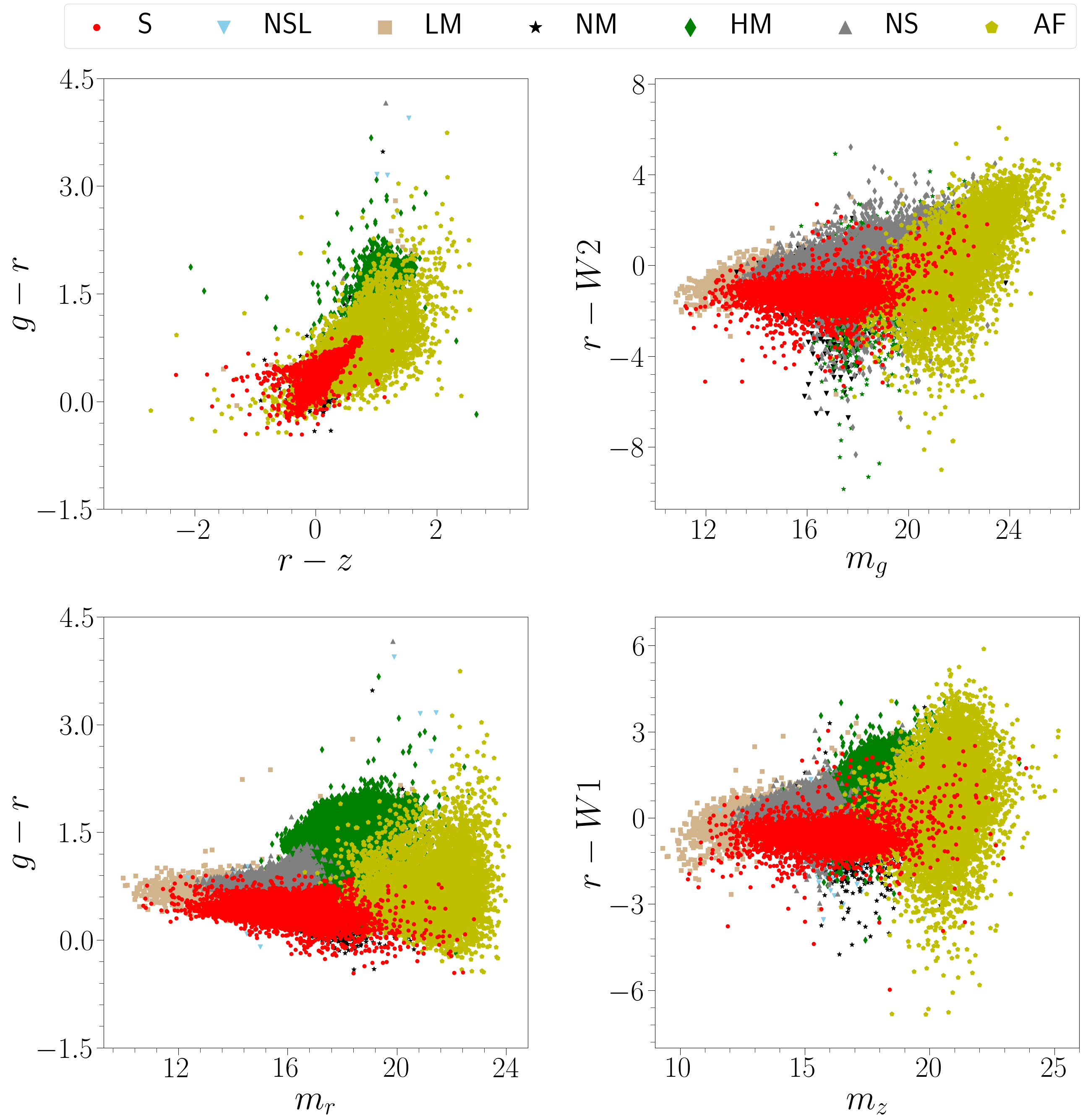}
\end{center}
\caption{The color-color distribution of the signal and contaminants in the training dataset. 
Red dot represents the signal (local dwarf galaxies), 
sky-blue inverted triangle represents nearby sub-$L^*$ galaxies,
tan square stands for local massive galaxies, 
black star denotes nearby massive galaxies, 
green diamond marks high-z massive galaxies, 
gray triangle represents nearby spiral galaxies 
and yellow pentagon stands for artifact Features.} 
\label{fig:colorcut}
\end{figure*}

We present a set of color-magnitude/color-color diagrams of the training sample in Figure \ref{fig:colorcut}. Clearly, no single feature (magnitude or color) is able to separate signal from the contaminants effectively. Obviously, a traditional two-dimensional, color-magnitude/color-color selection method is ineffective and results in low precision and low recall. However, we would expect that the machine learning algorithms would significantly enhance the performance in the separation between the signal and the contaminants when making full use of all magnitudes/colors or surface brightness information in the high-dimensional feature space. 

\subsection{Comparison of Different Algorithms} \label{sec:ComALgo}

There are many machine learning algorithms with their own strengths and weaknesses. Here we compare the performance of five different machine learning or deep learning algorithms for our particular aim of separating the local dwarf galaxies from various contaminants. The classification algorithms we examine in this study are the following:

\begin{enumerate}

\item \texttt{K-Nearest Neighbors (KNN)}: This is a non-parametric algorithm that forms a model by placing the input data in the high-dimensional feature space and making predictions based on the distance between the data point and its $k$ nearest neighbors. 
The main hyper-parameters in this algorithm include {\it the number of neighbors ($k$), weights ($w$)} and definition of {\it distance ($p$)}. The weights, for example, can be uniform or greater for closer  neighbors. The distance metric can be either the Manhattan or Euclidean distance.

\item \texttt{Random Forest (RF)}: This is an integrated approach that classifies by combining multiple decision tree models, where each tree is constructed using a random subset of the training data and a random subset of features. This randomness helps to reduce the overfitting issue and to improve generalisation. The main hyper-parameters of this algorithm include {\it the number of decision trees that make up the model, the criterion used to quantify the split quality, the maximum depth of the trees, the minimum number of samples required to split an internal node, the minimum number of samples required at each leaf node, the maximum number of features to consider for each split}. 

\item \texttt{XGBoost (eXtreme Gradient Boosting)}: This is a gradient-boosting decision tree algorithm designed to be highly efficient, flexible, and portable. It constructs a robust ensemble model by iteratively training decision tree models. In each iteration, a new tree is fitted to the residuals of the current model and added to the ensemble to enhance model performance. While XGBoost (XGB) is a form of boosting, it supports parallel computation for optimal feature search at tree nodes, significantly speeding up the training process. Additionally, the randomness in random forest algorithm is also implemented by XGBoost for random sampling of rows and columns. The addition of regularization terms in the objective function effectively helps to reduce model complexity to prevent overfitting. The main hyper-parameters of this algorithm include {\it all of the parameters of the decision tree algorithm, plus the number of estimators, the learning rate, the proportion of the training sample used, the number of parallel threads, the loss function, and the regularization parameters}.

\item \texttt{Neural Network \citep[NN,][]{McCulloch1990, Chollet2018}}: The network is a structure consisting of interconnected computational nodes, or `neurons', arranged in layers. It typically consists of multiple layers, including input, hidden, and output layers. Between neurons in adjacent layers, there exists a set of weighted sequences to transmit data. The backpropagation algorithm in neural networks updates the weights by calculating the gradients of the weights with respect to the loss function, aiming to make the network's predictions closer to the actual values. Additionally, neural networks can divide training samples into multiple batches to speed up the training process and to reduce the overfitting issue. The main hyper-parameters of this algorithm include {\it the learning rate, the optimization algorithms, the batch size, the number of epochs, the number of hidden units, the activation function, and the regularization parameters}.

\item \texttt{TabNet \citep[TB,][]{Arik2019}}: This approach is a new canonical deep neural network (DNN) architecture for tabular data, which trains on tabular data using gradient descent-based optimization. TabNet (TB) uses sequential attention to choose which features to reason from at each decision step and employs a single deep learning architecture for feature selection and reasoning. The main hyper-parameters of this algorithm include {\it the dimensions of the feature transformer and hidden units in the decision step, the number of attention units in the feature transformer, the number of decision steps, the batch size, the learning rate, the mask type, the choice of optimizer}.

\end{enumerate}

\begin{table*}[]
\centering
\begin{tabular}{|c|ccc|ccc|ccc|ccc|}
\hline
\multirow{3}{*}{}
& \multicolumn{3}{c|}{Precision}                             
& \multicolumn{3}{c|}{Recall}                                
& \multicolumn{3}{c|}{f1 score}                             
& \multicolumn{3}{c|}{AGF}               
\\ \cline{2-13} 
& \multicolumn{1}{c|}{Validation}                
& \multicolumn{2}{c|}{Test}            
& \multicolumn{1}{c|}{Validation}                
& \multicolumn{2}{c|}{Test}           
& \multicolumn{1}{c|}{Validation}                
& \multicolumn{2}{c|}{Test}            
& \multicolumn{1}{c|}{Validation}                
& \multicolumn{2}{c|}{Test}            
\\ \cline{2-13} 
& \multicolumn{1}{c|}{signal}            
& \multicolumn{1}{c|}{signal} & others 
& \multicolumn{1}{c|}{signal}             
& \multicolumn{1}{c|}{signal} & others 
& \multicolumn{1}{c|}{signal}            
& \multicolumn{1}{c|}{signal} & others 
& \multicolumn{1}{c|}{signal}            
& \multicolumn{1}{c|}{signal} & others 
\\ \hline

KNN 
& \multicolumn{1}{c|}{0.886\(\pm 0.010\)} & \multicolumn{1}{c|}{0.901}  & 0.917  
& \multicolumn{1}{c|}{0.772\(\pm 0.007\)} & \multicolumn{1}{c|}{0.786}  & 0.922  
& \multicolumn{1}{c|}{0.825\(\pm 0.007\)} & \multicolumn{1}{c|}{0.840}  & 0.919  
& \multicolumn{1}{c|}{0.897\(\pm 0.005\)} & \multicolumn{1}{c|}{0.896}  & 0.920  \\ \hline

RF
& \multicolumn{1}{c|}{0.892\(\pm 0.008\)} & \multicolumn{1}{c|}{0.914}  & 0.926  
& \multicolumn{1}{c|}{0.786\(\pm 0.010\)} & \multicolumn{1}{c|}{0.792}  & 0.932  
& \multicolumn{1}{c|}{0.835\(\pm 0.008\)} & \multicolumn{1}{c|}{0.848}  & 0.928  
& \multicolumn{1}{c|}{0.894\(\pm 0.006\)} & \multicolumn{1}{c|}{0.899}  & 0.956  \\ 
\hline

XGB 
& \multicolumn{1}{c|}{0.895\(\pm 0.003\)} & \multicolumn{1}{c|}{0.912}  & 0.936  
& \multicolumn{1}{c|}{0.826\(\pm 0.005\)} & \multicolumn{1}{c|}{0.833}  & 0.939  
& \multicolumn{1}{c|}{0.859\(\pm 0.002\)} & \multicolumn{1}{c|}{0.870}  & 0.937  
& \multicolumn{1}{c|}{0.913\(\pm 0.003\)} & \multicolumn{1}{c|}{0.918}  & 0.962  \\ 
\hline

NN 
& \multicolumn{1}{c|}{0.953\(\pm 0.012\)} & \multicolumn{1}{c|}{0.953}  & 0.936 
& \multicolumn{1}{c|}{0.740\(\pm 0.031\)} & \multicolumn{1}{c|}{0.761}  & 0.943  
& \multicolumn{1}{c|}{0.833\(\pm 0.016\)} & \multicolumn{1}{c|}{0.846}  & 0.939 
& \multicolumn{1}{c|}{0.877\(\pm 0.017\)} & \multicolumn{1}{c|}{0.887}  & 0.964  \\ 
\hline
TB
& \multicolumn{1}{c|}{0.934\(\pm 0.018\)} & \multicolumn{1}{c|}{0.948}  & 0.924  
& \multicolumn{1}{c|}{0.743\(\pm 0.038\)} & \multicolumn{1}{c|}{0.754}  & 0.930  
& \multicolumn{1}{c|}{0.827\(\pm 0.020\)} & \multicolumn{1}{c|}{0.840}  & 0.926   
& \multicolumn{1}{c|}{0.877\(\pm 0.020\)} & \multicolumn{1}{c|}{0.883}  & 0.954  \\
\hline
\end{tabular}
\caption{\label{tab:algorithms}Evaluation scores for the five different classification algorithms on the validation set and the test set. In addition, we present the average of the metrics weighted by the number of instances in each class for the contaminants (``others") for the test set.}
\end{table*}

We apply the above algorithms to train a multi-class (7 classes as shown in Table \ref{tab:training}) classification model and obtain the corresponding precision, recall, f1 score and AGF score for each of these models in classifying local dwarf galaxies (Table~\ref{tab:algorithms}). We also present the weighted average of each metric, weighted by the number of instances in each class, for all of the contaminants, in the test set. For model training, we use {\it RandomizedSearchCV} to find the best set of hyperparameters in the model's hyperparameter space. We use cross-validation (CV) to avoid overfitting on the training sample, and the fold number is set to be 5. We adopt the standard deviation within the five folds as the error. The results from the validation set (part of the training sample) is indicated by ``Validation" column in Table \ref{tab:training}. 

As we show in Table \ref{tab:algorithms}, all five algorithms perform quite well, which indicates that the signal indeed possesses intrinsic characteristics that make it separable from overwhelming contaminants. Overall,  although the recall of the neural network is the lowest among all of the algorithms, the neural network reaches the highest precision and XGB the highest recall, f1 and AGF scores. As discussed earlier, we choose to emphasize precision, so we select to proceed with the neural network algorithm. 

In Table~\ref{tab:FI_nn}, we list the top 20 features, ranked by importance of the neural network classification. As is clearly seen, the apflux ratios are critical in the neural network classification model, which demonstrates that the surface brightness profile is key in separating the signal from the otherwise overwhelming number of contaminants. 
In Table~\ref{tab:NN_model}, we present the architecture of our final NN model.



\begin{table}
\centering
\begin{tabular}{lc}
\toprule
Feature   & Importance [\%]  \\
\toprule


\text{apflux\_g\_6}/\text{apflux\_W1\_1} & 9.292 \\
\text{r}-\text{z} & 8.109 \\
\text{apflux\_g\_7}/\text{apflux\_W1\_2} & 6.084 \\
\text{apflux\_r\_7}/\text{apflux\_W1\_2} & 5.383 \\
\text{apflux\_r\_6}/\text{apflux\_W1\_1} & 5.367 \\
\text{apflux\_g\_6}/\text{apflux\_r\_6} & 5.316 \\
\text{apflux\_r\_8}/\text{apflux\_W1\_3} & 4.562 \\
\text{apflux\_g\_3}/\text{apflux\_z\_3} & 3.947 \\
\text{apflux\_g\_2}/\text{apflux\_z\_2} & 3.609 \\
\text{apflux\_r\_8}/\text{apflux\_W2\_3} & 3.141 \\
\text{g}-\text{r} & 3.076 \\
\text{apflux\_g\_4}/\text{apflux\_z\_4} & 3.002 \\
\text{apflux\_z\_5}/\text{apflux\_z\_6} & 2.948 \\
\text{apflux\_g\_8}/\text{apflux\_z\_8} & 2.725 \\
\text{apflux\_z\_6}/\text{apflux\_z\_7} & 2.682 \\
\text{apflux\_g\_7}/\text{apflux\_r\_7} & 2.512 \\
\text{apflux\_g\_2}/\text{apflux\_g\_3} & 2.396 \\
\text{apflux\_r\_5}/\text{apflux\_r\_6} & 2.348 \\
\text{apflux\_W1\_2}/\text{apflux\_W1\_3} & 2.345 \\
\text{apflux\_z\_4}/\text{apflux\_z\_5} & 2.117 \\

		\toprule
	\end{tabular}
	\caption{\label{tab:FI_nn}The top 20 feature in terms of their permutation feature importance ranking provided by the neural network classification model.} 
\end{table}

\begin{table}[h!]
\centering
\begin{tabular}{|c|c|c|c|}
\hline
Layer & Configuration & Activation & Dropout \\ \hline
Input    & Flatten    & - & - \\ \hline
Hidden 1 & Dense(64)  & ReLU & 0.2 \\ \hline
Hidden 2 & Dense(256) & ReLU & 0.2 \\ \hline
Hidden 3 & Dense(32)  & ReLU & 0.2 \\ \hline
Output   & Dense(7)   & Softmax & - \\ \hline
\end{tabular}
\caption{The architecture of our final neural network model.}
\label{tab:NN_model}
\end{table}

\subsection{Confusion Matrix} \label{sec:CM}

\begin{figure*}[ht]
\begin{center}
\includegraphics[width = 0.875 \textwidth]{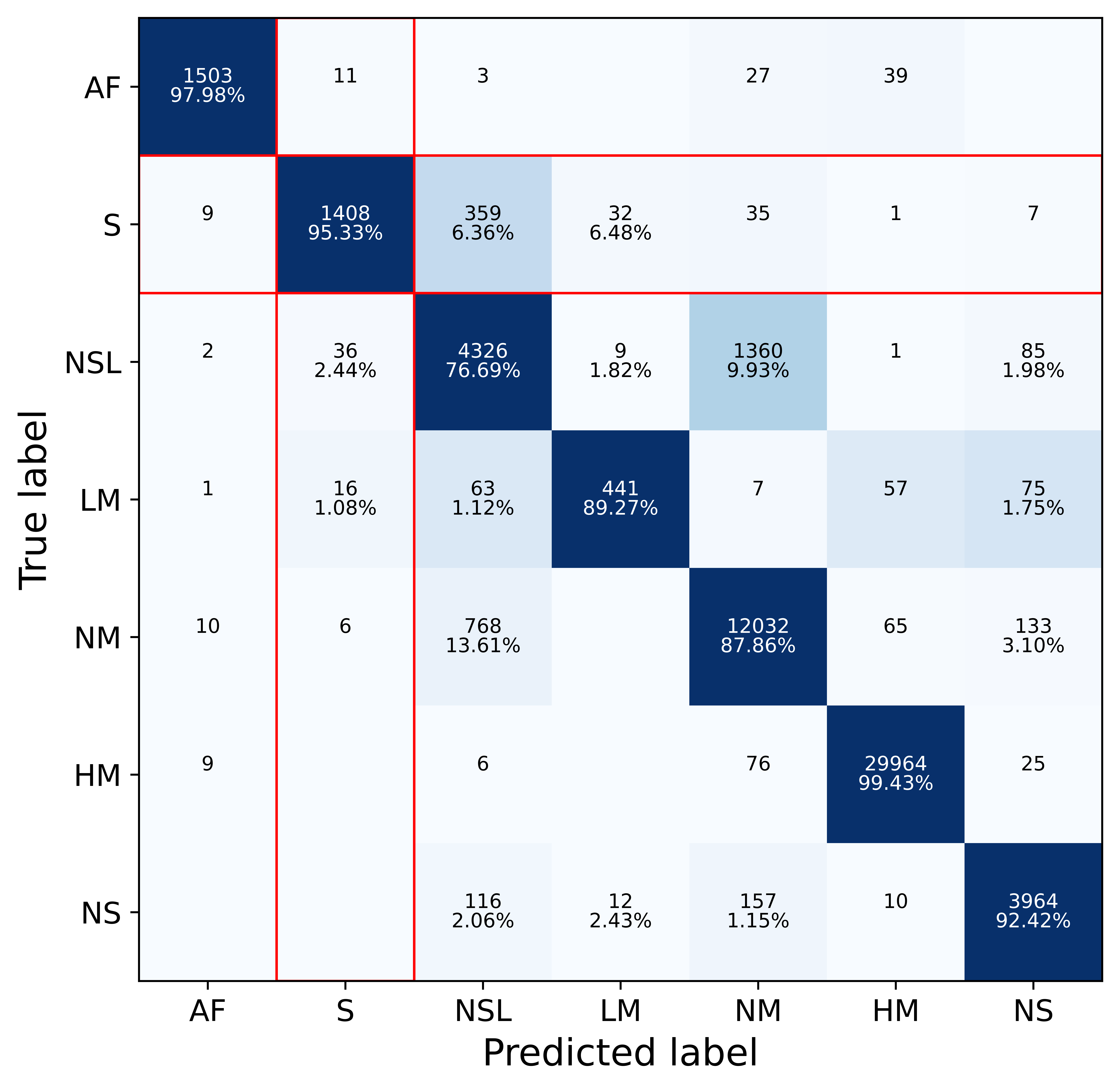}
\end{center}
\caption{Confusion matrix of the 7-class neural network classification model for the test set. Each column in each row represents the predicted class counts for the corresponding true class. The percentages on the diagonal represent the precision of each class, and the shades of color represent the recall of each class. The precision of the signal is $95.33\%$, and the recall of the signal can be calculated simply to be is $76.07\%$, which means that more than $20\%$ of the signal are incorrectly predicted as other types. }
\label{fig:CM}
\end{figure*}

In Figure \ref{fig:CM} we show the neural network confusion matrix for the test set, where each row represents the true class and each column represents the predicted class. The percentages in each square indicate the fraction of the corresponding class in each column. Squares with a percentage $<$ 1\% are not displayed. Along the diagonal, the percentages represent the precision of the model for the corresponding class. 
For our signal, local dwarf galaxies, the precision is $95.33\%$ and the recall is $76.07\%$ for the test set. 



The principal contaminants among the local dwarf galaxy sample are nearby sub-$L^*$ galaxies, local massive galaxies and nearby massive galaxies. These classes all have members that are similar to our signal class in terms of redshift and stellar mass, resulting in similar colors and surface brightness profiles. Surprisingly, we also find a number of artifacts that are classified as signal, which indicates that images of higher quality would provide further gains. 

\begin{figure*}[ht]
\begin{center}
\includegraphics[width = \textwidth]{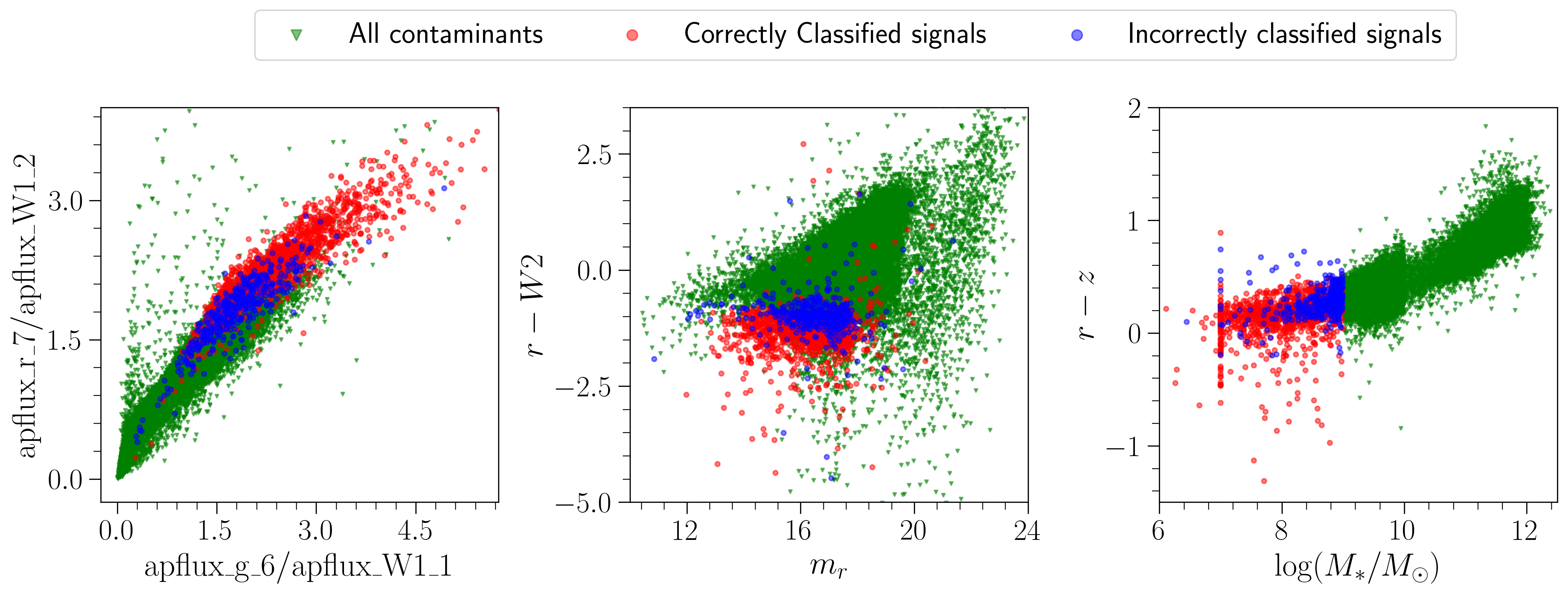}
\end{center}
\caption{The distribution of the contamination and the signal for the test set in the parameter space. Green triangles represent all contaminants, red points indicate correctly classified signals, and blue points stand for misclassified signals.}
\label{fig:result_test}
\end{figure*}

The relatively low recall on the test set indicates that a considerable portion of the signal was not correctly identified by our model. In Figure \ref{fig:result_test}, we present the distribution of the signal and the contamination in the parameter space. 
One can easily observe that the misclassified signals are much closer to the contaminators than the correctly classified signals, 
which illlustrates graphically the limitation 
of our classification model. 

\section{Candidates and Verification} \label{sec:candidates}

As noted, we adopt the 7-class neural network model to search for local dwarf galaxies. To do so, we will apply the trained classification model to the entire Legacy Survey DR9 dataset. Before doing so, however, we impose several selection criteria to reduce the size of the LS DR9 catalog data (over 1 billion entries) and to obtain more reliable results. Our selection criteria, 
followed by our reasoning, are as follows: 
\begin{enumerate}
    \item $dered\_mag\_g,r,z,W1,W2$ is not null.
     The trained classification model cannot not work with missing values. 
     
    \item half light radius greater than 1 arcsecond. Due to the survey depth and resolution, such small objects are mostly too faint and their photometry is unreliable. 

    \item $brick\_primary=1$, 
    $maskbits=0$, $wisemask\_{\it W1, W2}=0$,
    and $anymask\_{\it g, r, z}=0$. These are a set of criteria that ensure measurement quality. $brick\_primary=1$ ensures that the sources lie entirely within a given brick. $maskbits=0$ ensures that sources are pure and uncontaminated.
    $wisemask\_W1,W2=0$ ensures that sources are also not contaminated in the $W1, W2$ bands. 
    $anymask\_{g, r, z}=0$ ensures that the source does not touch any bad pixels in all of a set of overlapping {\it g, r, z} band images. 
    
    \item $snr\_g>5$, $snr\_r>5$.
    This simply removes objects with less reliable photometry. 
    \item $dered\_mag\_r<21.5$. This criterion removes faint objects that are also likely to have unreliable photometry and that will also be more challenging for spectroscopic follow-up. 
\end{enumerate}

The number of samples obtained by incrementally adding selection criteria is presented in Table \ref{tab:cuts}.

\begin{table}
\centering
\begin{tabular}{|c|c|}
\hline
Selection Criteria & Numbers \\ \hline
All sky 
& 1,969,942,678 \\ 
\hline
NOT NULL 
& 1,067,722,306 \\ 
\hline
$r_{1/2} > 1^{\prime\prime}$ 
& 94,874,076 \\ 
\hline
\parbox{4.4cm}{\centering Brick Primary = 1, \\ Maskbits = 0,\\ Wisemask W1, W2 = 0,\\ Anymask g, r, z = 0} 
& 85,333,687 \\ 
\hline
\parbox{2.2cm}{\centering $\text{snr g} > 5$,\\ $\text{snr r} > 5$} & 71,126,585 \\ 
\hline
$\text{mag\_r} < 21.5$ 
& 39,820,031 \\ 
\hline
NN classifier, $\text{mag\_z} \leq 22.5$
& 112,859 \\ 
\hline
\end{tabular}
\caption{\label{tab:cuts} The number of samples obtained by incrementally adding selection criteria.}
\end{table}



\subsection{Local dwarf galaxy candidates}

After applying the above selection criteria, there are 39,819,821
sources from the entire LS DR9 to pass to the trained neural network classification model. For each source, the neural network classification algorithm assigns a probability that the source belongs to each of the 7 allowed classes. We classify the object by placing it in the class for which it has the highest predicted probability. This choice results in some candidates having a low probability of being in the selected class. Our choice is to present all of the candidates. Anyone using the catalog can impose a floor probability of their choice. We present the number of objects in each predicted class in Table \ref{tab:pred}.

We classify 113,616 sources as local dwarf galaxies.  We notice that even though our selection criteria require $mag\_r < 21.5$ to remove faint objects, our signal candidates still include very faint sources which are almost invisible in the images. In our investigation of these, we find that these faint sources are contaminated by strong starlight. Considering that the Legacy Survey \citep{Dey2019} reaches $5\sigma$ depths of $g = 24.0, r = 23.4$, and $z = 22.5$ AB mag for faint galaxies, we remove signal candidates with $mag\_z > 22.5$, totaling 757 (for $mag\_z > 23$: 491), most of which are not visible in the images. We retain 112,859 local dwarf galaxy candidates.


\begin{table*}[t]
\centering
\begin{tabular}{|c|c|}
\hline
Category & Number  
\\ \hline
Signal ($z < 0.02$ and $10^6 < M_* < 10^9$ M$_\odot$)  
& 112,859  
\\ \hline
Nearby Sub-$L^*$ ($0.02 < z < 0.05$ and $10^{9} < M_* < 10^{9.75}$ M$_\odot$) 
& 99,434
\\ \hline
Local Massive ($z < 0.02$ and $M_* > 10^{9.5}$ M$_\odot$)
& 15,610  
\\ \hline		
Nearby Massive ($0.05 < z < 0.2$ and $10^{9.25} < M_* < 10^{10}$ M$_\odot$) 
& 6,618,376 
\\ \hline
High-z Massive  ($z > 0.2$ and  $M_* > 10^{10}$ M$_\odot$)  
& 25,492,266
\\ \hline
Nearby Spiral ($z > 0.03$)    
& 417,185
\\ \hline
Artifact Features     
& 7,063,334
\\ \hline											
\end{tabular}
\caption{\label{tab:pred}  The number of objects in each category in the entire LS DR9 footprint. We take the class with the highest predicted probability as the predicted category for each source object. For our signal candidates, the numbers represent the outcomes following the exclusion of faint sources. } 
\end{table*}

\begin{figure}[ht]
\begin{center}
\includegraphics[width = 0.48 \textwidth]{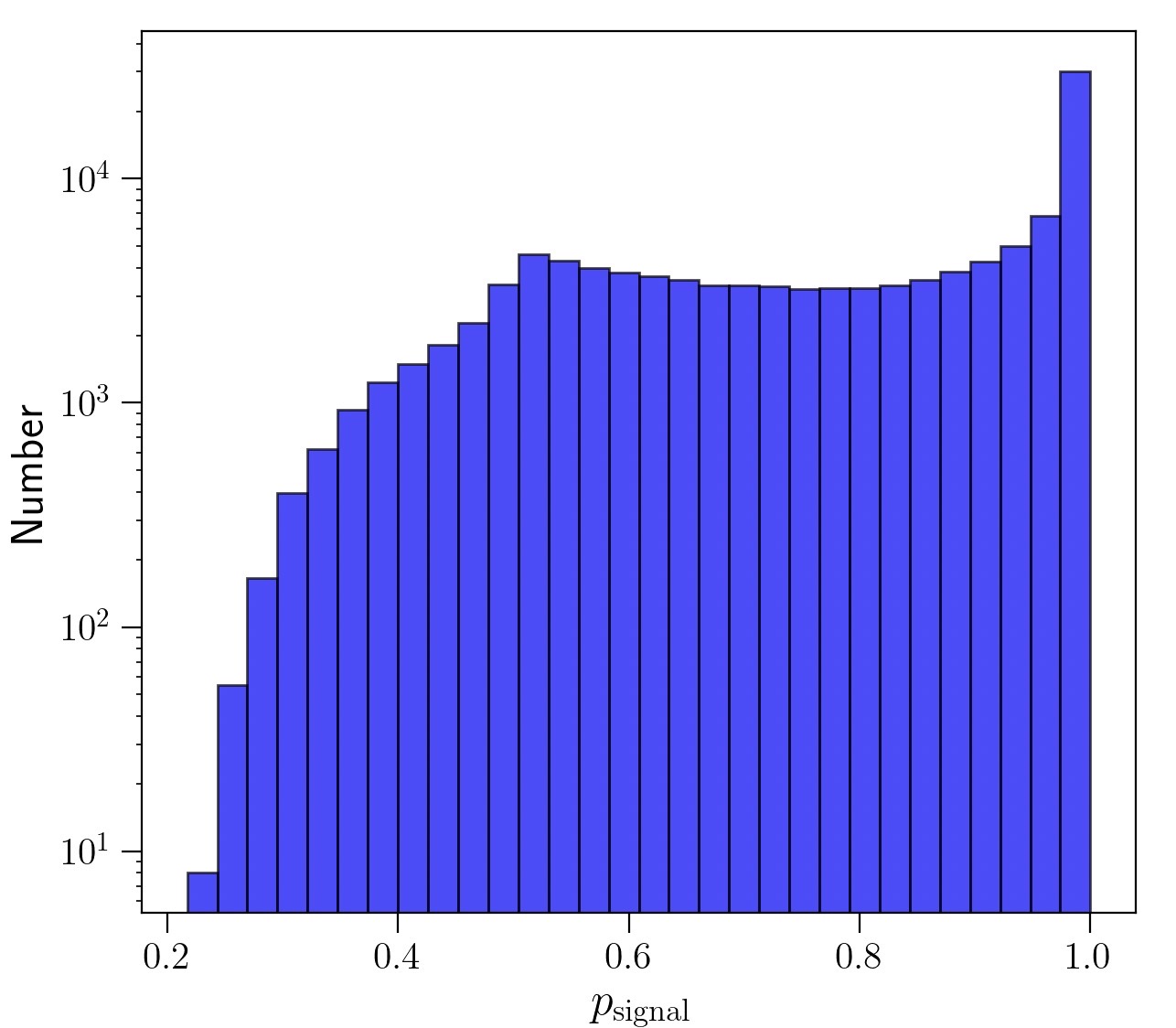}
\end{center}
\caption{The probability distribution of the local low-mass dwarf galaxy candidates provided by the neural network model.}
\label{fig:candidate_prob}
\end{figure}

We present the distribution of probabilities for the sources classified as local dwarf galaxy candidates in Figure \ref{fig:candidate_prob}. The number of candidates per probability bin increases steeply as the probability increases from 0.2 to 0.5, and flattens in the probability range of $0.5 < p_{\rm signal} < 0.95$. The fraction of candidates with $p_{\rm signal} < 0.5$ is 10.25\%, which is much higher than that of the test set (1.90\%) in the training sample. So, candidates with $p_{\rm signal} < 0.5$ may not be reliable. Surprisingly, the numbers increase sharply again for $p_{\rm signal} > 0.95$. There are 45,460, 36,408 and 23,370 local dwarf galaxy candidates with $p > 0.9, 0.95$ and $0.99$, respectively. Selecting a high-confidence sample for spectroscopic follow-up, regardless of the exact choice of $p$ threshold within this range, will yield tens of thousands of candidates. 


\begin{figure}[ht]
\begin{center}
\includegraphics[width = 0.48 \textwidth]{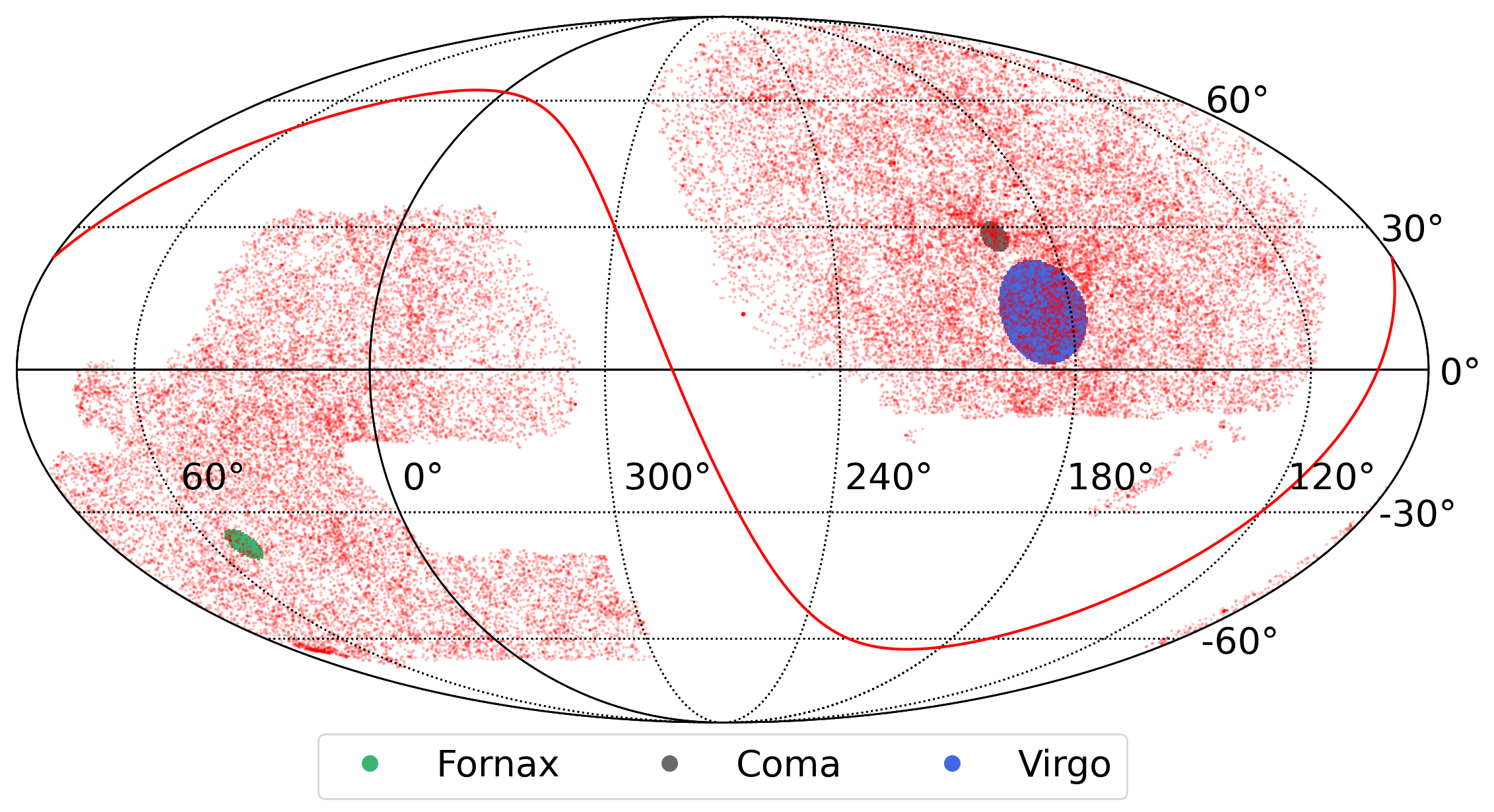}
\includegraphics[width = 0.48 \textwidth]{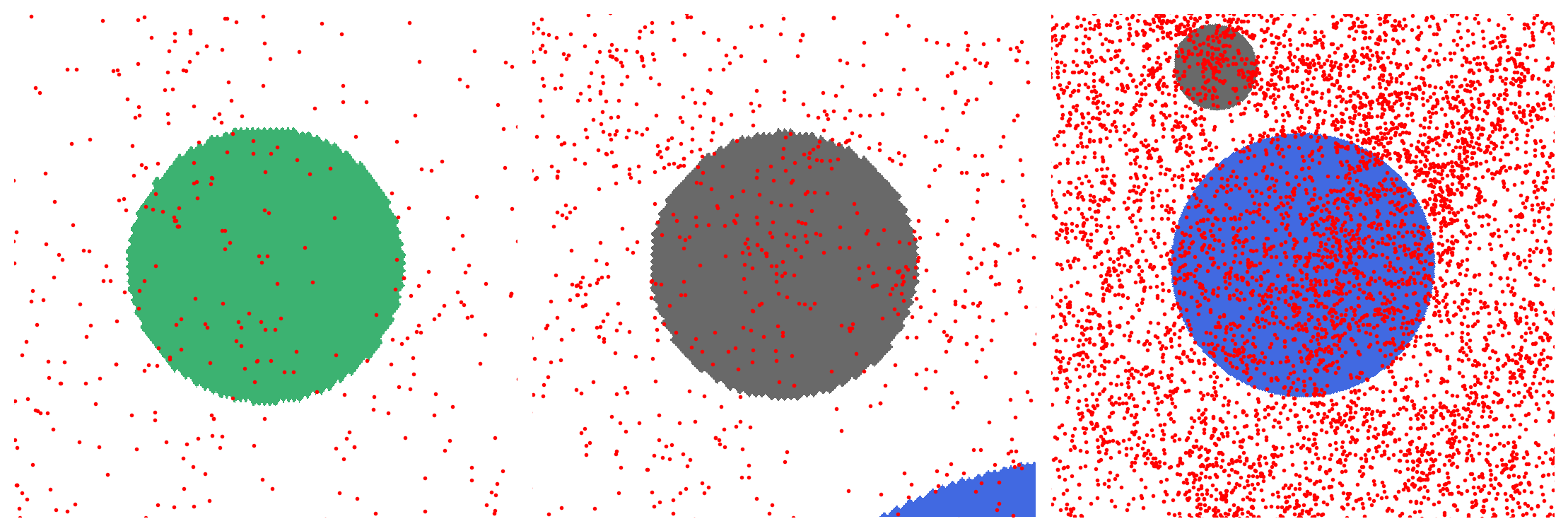}
\end{center}
\caption{The distribution of the local low-mass  galaxy candidates ($p_{\rm signal} > 0.90$) across the sky in R.A. and decl (top) and the zoom-in for the local galaxy clusters (bottom). We label the well-known local over-densities (Fornax, Coma, and Virgo) for reference. The sizes of the symbols reflect twice of virial radius of those over-densities. The virial radius of those over-densities are as follows, Fornax: 0.49 Mpc or 1.71$^\circ$ from \cite{Maddox2019}; Coma: 1.99 Mpc or 1.66$^\circ$ from \cite{Kubo2007}; Virgo: 1.08 Mpc or 5.50$^\circ$ from \cite{Urban2011}. }
\label{fig:coord}
\end{figure}

We present the on-sky distribution of the local dwarf galaxy candidates with probability greater than 0.90 in Figure \ref{fig:coord}. Additionally, we also assign those candidates
projected onto the Coma, Fornax, and Virgo clusters, the richest clusters in the nearby Universe, to those environments. Somewhat surprisingly, we do not find strong enhancements of candidates superposed on the Coma, Fornax, and Virgo clusters. This may, in part at least, reflect our exclusion of objects whose photometry may be contaminated by nearby neighbors.

\begin{figure*}[ht]
\begin{center}
\includegraphics[width = 0.8 \textwidth]{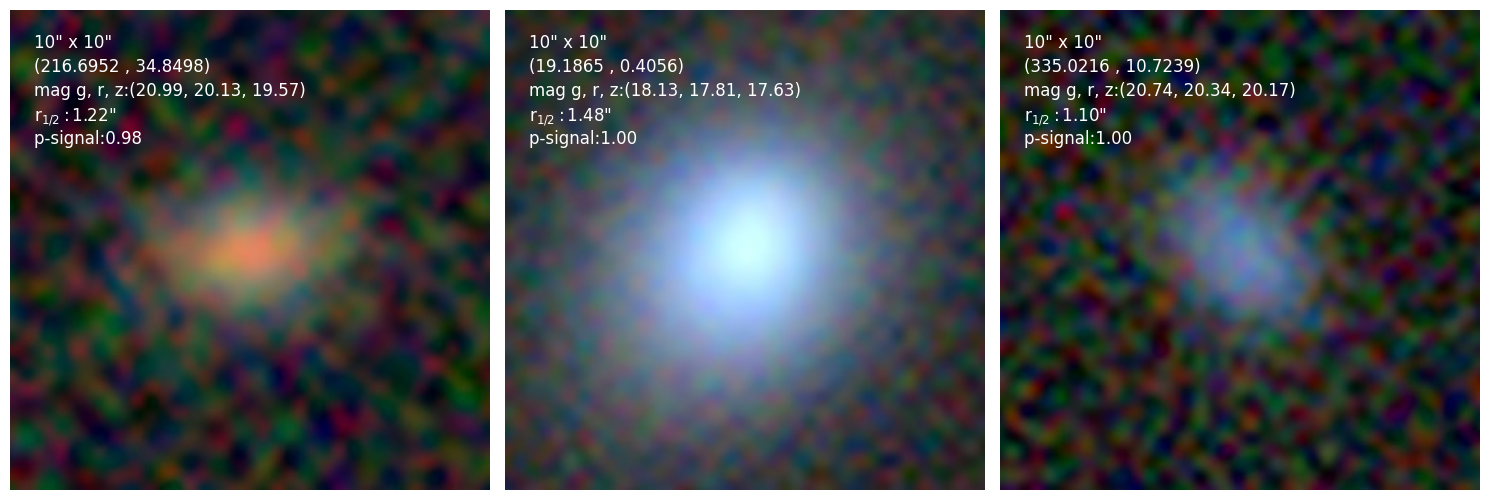}\\
\includegraphics[width = 0.8 \textwidth]{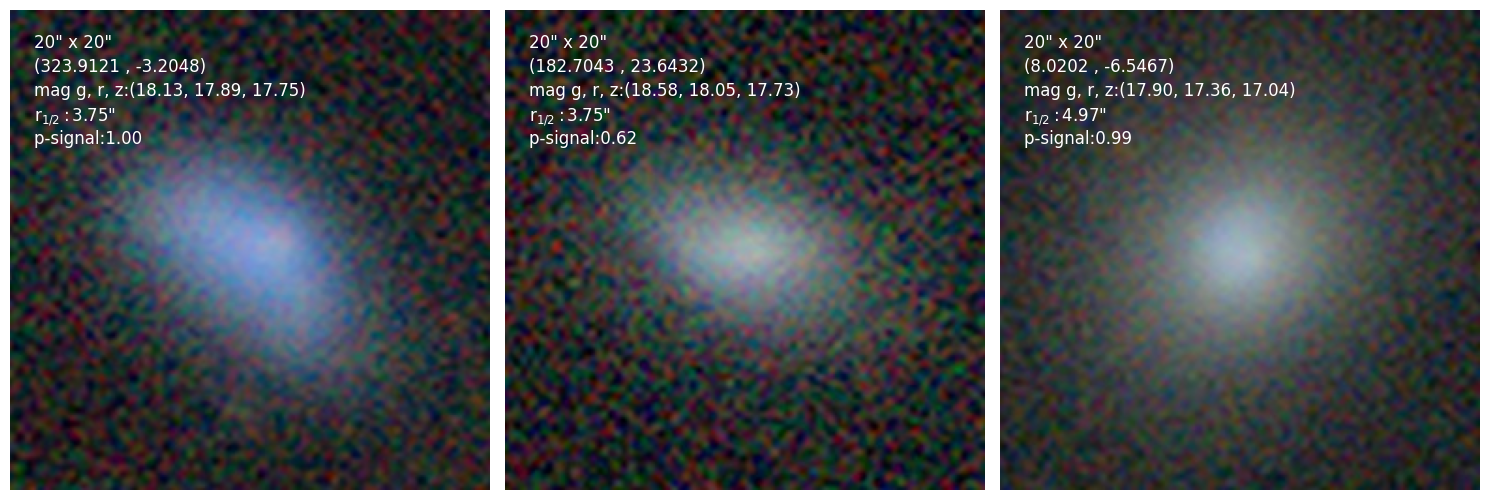}\\
\includegraphics[width = 0.8 \textwidth]{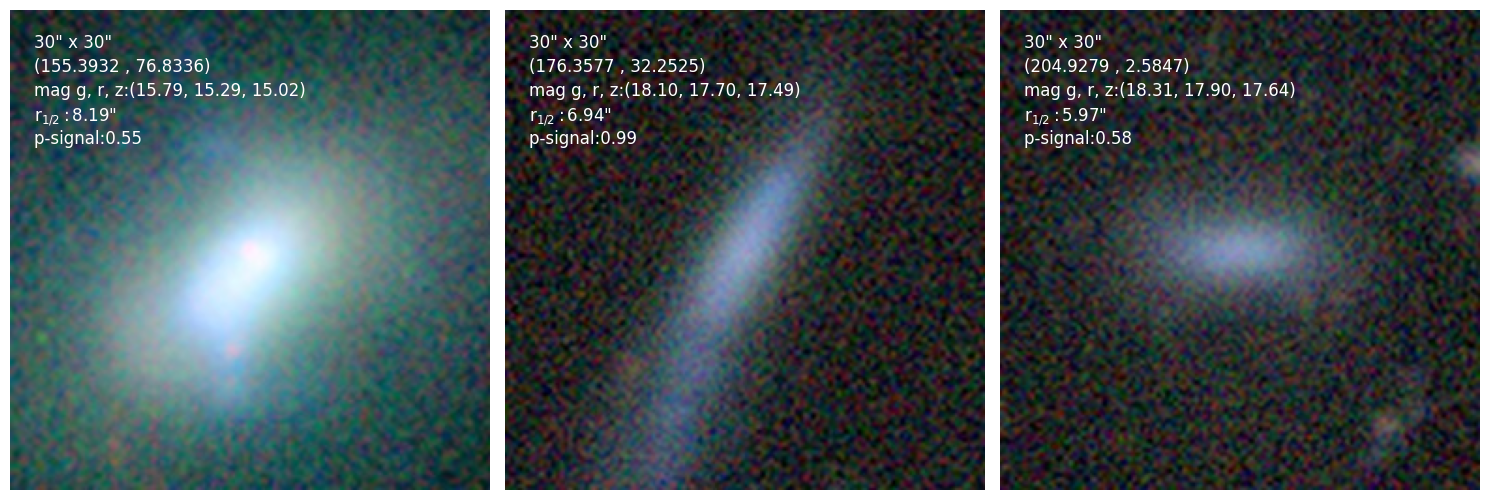}
\end{center}
\caption{Cutout images of some local dwarf galaxy candidates. The majority of those candidates show very blue colors and the structures are diverse.}
\label{fig:candcutout}
\end{figure*}

\begin{figure*}[ht]
\begin{center}
\includegraphics[width = 0.8 \textwidth]{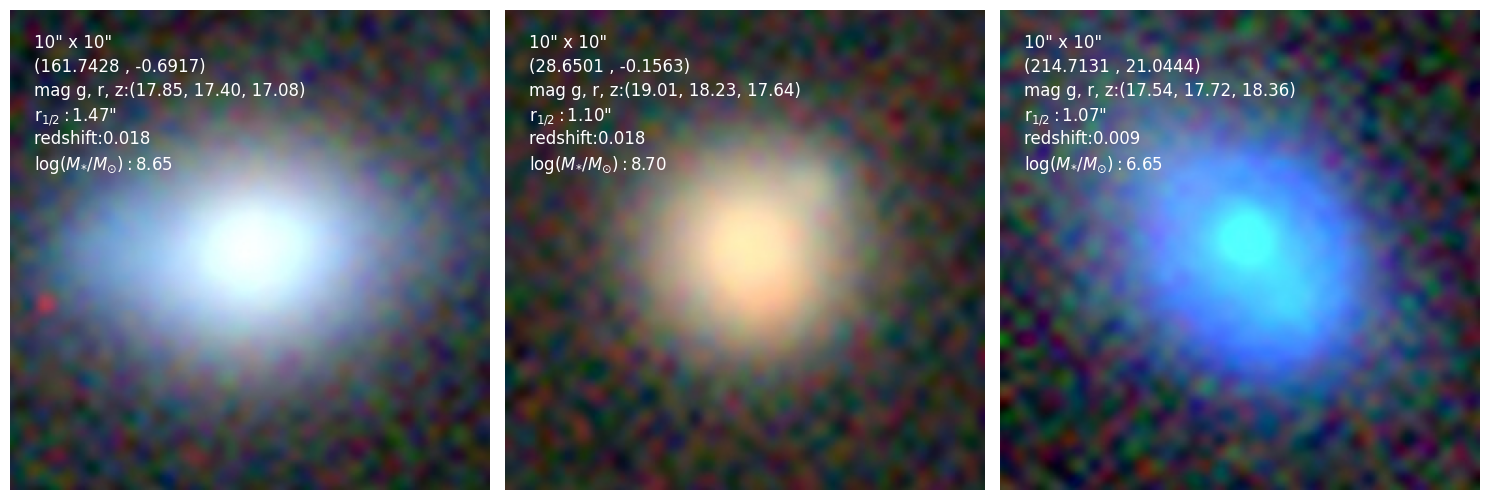}\\
\includegraphics[width = 0.8 \textwidth]{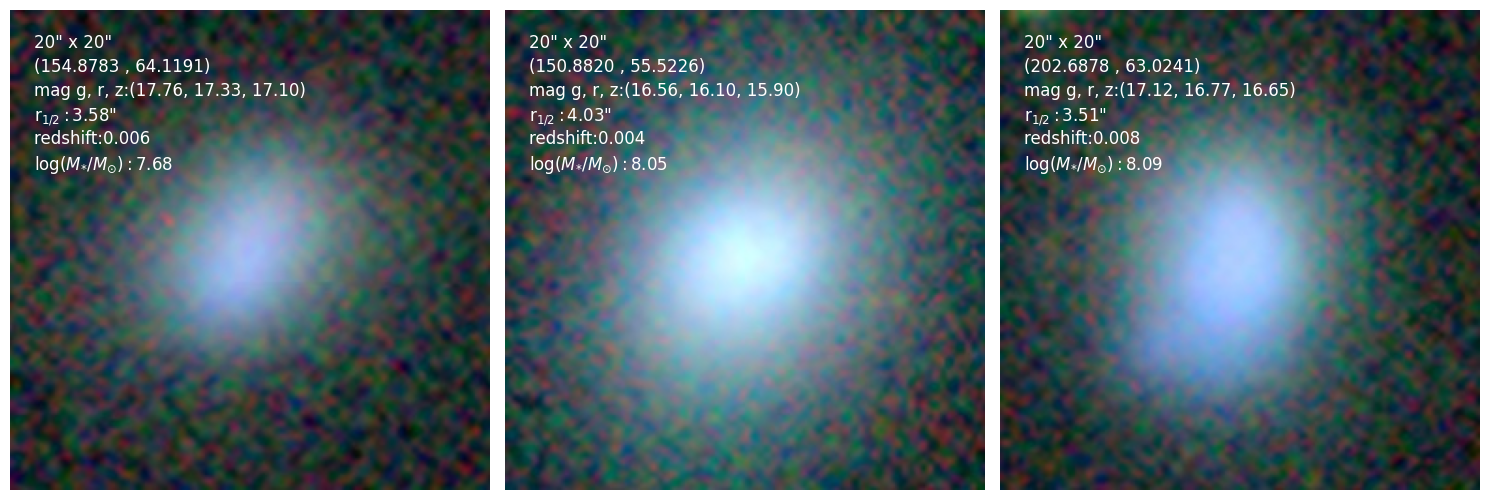}\\
\includegraphics[width = 0.8 \textwidth]{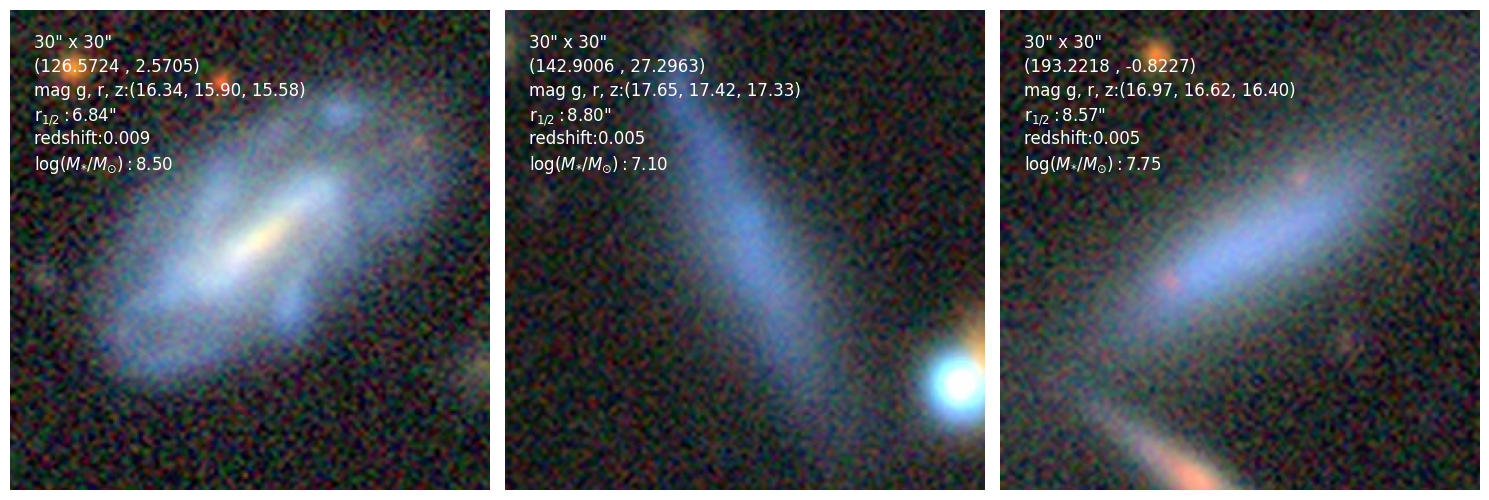}
\end{center}
\caption{Cutout images of some signal selected from the training sample. Surprisingly, 
the panel in the bottom left shows very obvious spiral arms and the panel in the top left has a very bright bulge.} 
\label{fig:signalcutout}
\end{figure*}

We present images of some of the local dwarf galaxy candidates as well as some true signal drawn from the training sample in Figures \ref{fig:candcutout} and \ref{fig:signalcutout}, respectively. The local dwarf galaxy candidates and the true dwarf galaxies in the training sample are qualitatively similar in terms of their color and structure. In Table \ref{tab:catalog} we present descriptions of the column entries in our catalog of local dwarf galaxy candidates, which is available in its entirety in machine-readable form. In Table \ref{tab:example}, we present some examples of the local dwarf galaxy candidates.

\begin{table}
	\centering
	\begin{tabular}{lr}
		\toprule
		Column name                             & {Description } \\
		\toprule
		RA               & R.A. in LS catalog   \\
		DEC     	     & Decl. in LS catalog   \\
		mag\_g    	     & {\it g} band magnitude in LS catalog   \\
		mag\_r    	     & {\it r} band magnitude in LS catalog   \\
 	      mag\_z    	   & {\it z} band magnitude in LS catalog   \\
            mag\_W1    	     & {\it W1} band magnitude in LS catalog   \\  
       	mag\_W2    	     & {\it W2} band magnitude in LS catalog   \\
            Prob   	        & Probability predicted to be the signal \\
            $r_{50}$        & half light radius in unit of arcsecond \\

   \toprule
	\end{tabular}

\caption{\label{tab:catalog} List of columns of the dataset consisting of the local low-mass galaxy candidates.}
\end{table}

\begin{table*}
\centering
\begin{tabular}{lllllllll}
\hline
\multicolumn{1}{c}{RA}      & \multicolumn{1}{c}{DEC}   
& \multicolumn{1}{c}{mag g} 
& \multicolumn{1}{c}{mag r} 
& \multicolumn{1}{c}{mag z} 
& \multicolumn{1}{c}{mag W1} 
& \multicolumn{1}{c}{mag W2} 
& \multicolumn{1}{c}{Prob.} & \multicolumn{1}{c}{$r_{50}$} \\
\multicolumn{1}{c}{(deg)}   & \multicolumn{1}{c}{(deg)} 
& \multicolumn{1}{c}{(AB)}  
& \multicolumn{1}{c}{(AB)}  
& \multicolumn{1}{c}{(AB)}  
& \multicolumn{1}{c}{(AB)}  
& \multicolumn{1}{c}{(AB)}   
& \multicolumn{1}{c}{}      & \multicolumn{1}{c}{(arcsec)} \\ 
\hline


119.6118                & ~10.4965
& 18.568                & 18.267           & 18.136
& 18.992                & 19.278
& 1.000                 & ~2.155
\\
174.0768                & ~25.3369
& 19.305                & 19.026           & 18.867
& 19.864                & 21.360
& 0.957                 & ~6.205
\\
140.4122                & ~28.1474
& 16.773                & 16.436           & 16.270
& 17.003                & 17.409
& 0.917                 & ~5.615                
\\
...                       & ...                 & ...     
& ...                     & ...                       
& ...                     & ...                     
& ...                     & ...                \\
~91.2223                & -48.9007
& 20.255                & 19.984           & 19.842
& 20.850                & 22.270
& 0.624                 & ~2.623
\\
~~0.9485                & ~~6.6186
& 20.667                & 20.281           & 20.092
& 18.929                & 19.272
& 0.582                 & ~2.525
\\
~~9.1897                & ~-9.3324
& 17.125                & 16.572           & 16.238
& 17.350                & 17.961
& 0.525                 & 13.414                      
\\
\hline
\end{tabular}
\caption{\label{tab:example} Some examples of our local dwarf galaxy candidates for $p_{\rm signal} >$ 0.5. Details see the machine readable table.}
\end{table*}

\subsection{Verification using DESI-EDR}

DESI will collect spectra for tens of millions of galaxies in its 5-year operation \citep{DESI1, DESI2, DESIover, DESIvalid}, including for many dwarf galaxies. There are 2,605 galaxies at $0.003 < z < 0.02$ in the DESI-EDR (Early Data Release) catalog \citep{DESI-EDR}. 
To see which of these are dwarf galaxies, we adopt the stellar mass estimates from the extensive catalog of \cite{Siudek2024}, where physical properties, such as stellar masses and star formation rates, were derived via spectral energy distribution fitting. 
1,869 out of 2,605 galaxies have the estimation of the stellar mass and star formation rate as extracted from the extensive catalog of \cite{Siudek2024}. 

Removing galaxies with missing values among any of our feature set in the DESI-EDR database, we are left with 1,383 galaxies for model validation, comprising of 1,255 dwarf galaxies and 128 local massive galaxies, with only 354 of those which are dwarfs being in our training sample. 
Because most of the galaxies at $z < 0.02$ in the DESI-EDR database are not present in the training sample, the galaxy ensemble in the DESI-EDR database serves as an excellent validation data set to test our model performance. 
Our model has the following outcomes: 640 are predicted to be signal, 212, 210, 146, 119, 37, 19 of them are classified as NM, NSL, AF, LM, HM, NS.
Of the 640 positive predictions, all 640 are true signals. 
The precision and recall for the signal using the independent DESI-EDR spectroscopic data is 100\% (640/640) and 51.00\% (640/1255), respectively. These results validate our emphasis on precision.  





\begin{figure*}[ht]
\begin{center}
\includegraphics[width = 0.875 \textwidth]{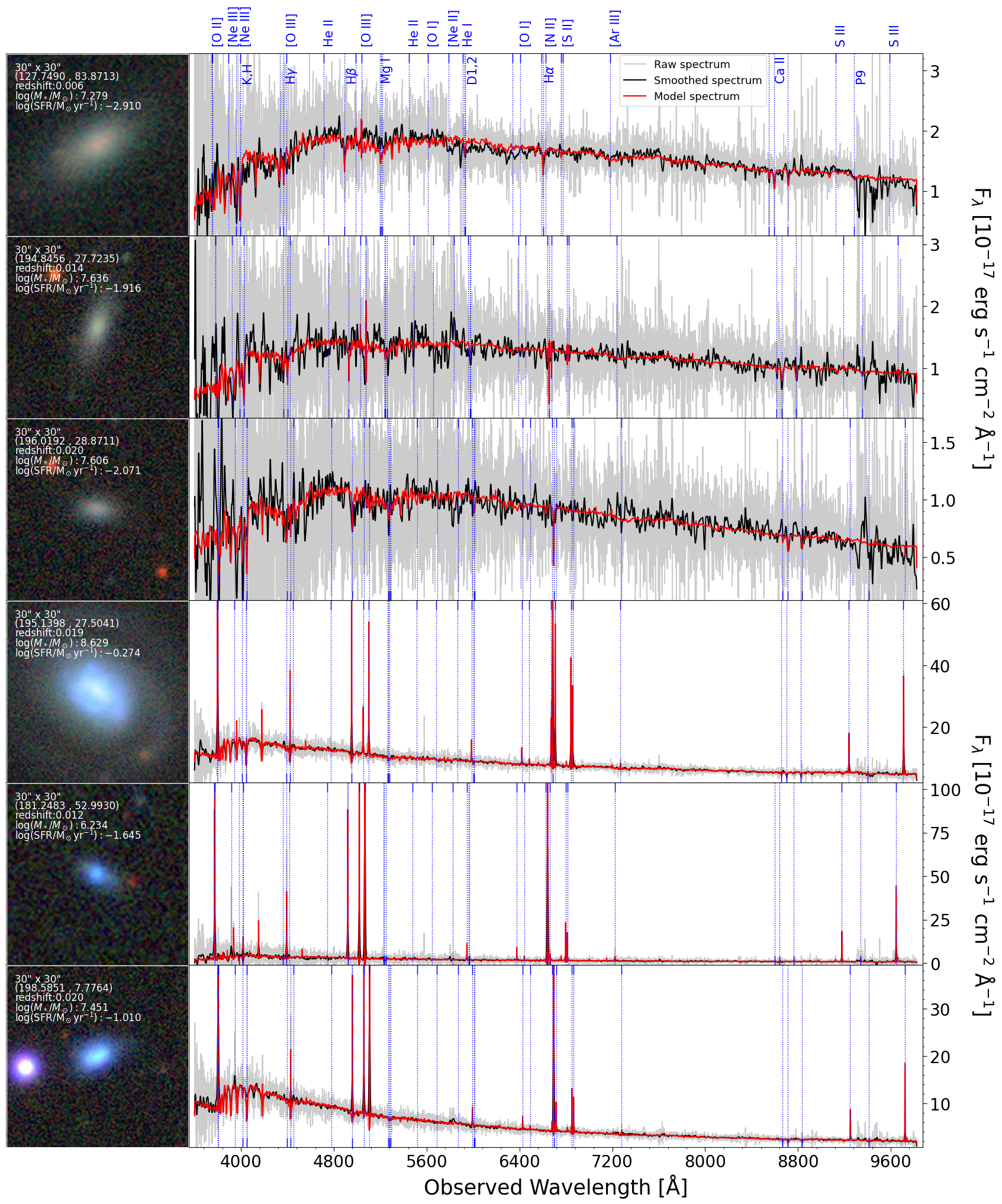}
\end{center}
\caption{The images and spectra of six signal galaxy candidates from DESI-EDR. The top three galaxies are comparably red, and their spectra do not show clear emission lines and are dominated by absorption lines. The bottom three galaxies are star-bursting galaxies, and their spectra are dominated by emission lines although the continuum is very low. The gray, black and red spectra represent the raw, smoothed, and model spectra, with the raw and model spectra sourced from DESI-EDR. The smoothed spectra are obtained by convolving the raw spectra with a one dimensional Gaussian, whose standard deviation is set to be 5. The vertical dashed lines indicate the position of featured emission lines and absorption lines.}
\label{fig:spec_edr}
\end{figure*}

In Figure \ref{fig:spec_edr} we present images and spectra for a selected group of 6 galaxies out of the DESI-EDR dwarf galaxy sample, which are confirmed to be signal. Three of these are comparably red and the spectra show stellar absorption features, while the other three are comparably blue and show strong emission features in the spectra. The classification model is able to identify both star-forming and quiescent dwarf galaxies. 

\subsection{Verification using SAGA}

As introduced in \S\ref{sec:intro}, the SAGA survey aims to measure the distribution of satellite galaxies ($M_r <-$ 12.3 mag, $M_* \gtrsim 3\times 10^6$ M$_\odot$) around 100 Milky Way analogs, which is defined based on $K-$band luminosity and the local environment \citep{SAGA2017}, in the distance range of 20 $-$ 40 Mpc ($0.005 < z < 0.01$). 
Their Stage II outcomes significantly enhance the sample size, containing 127 satellites with stellar mass estimations around 36 Milky Way analogs \citep{Mao2021}, with an enhanced target selection strategy and deep photometric imaging catalogs from DES and LS surveys. Therefore, all of the SAGA dwarf satellites are within the footprint of the LS. Because the SAGA data ensemble is not part of our training sample, it provides another independent test of our neural network classification model.

Among their 127 satellites, 118 have no missing feature values. Of those, 105 have a stellar mass in the range of $10^6$ M$_\odot$ to $10^9$ M$_\odot$ as given by SAGA, and so within the range covered by signals in our training sample. After applying our classification model, 74 out of all 118 satellites
are predicted to be signal and 71 of those are correctly identified. The major contamination to the signal is AF, which is almost identical to that for the training sample. The true labels of the three misclassifications ($z < 0.01$)  are LM, with predicted probability of 0.84, 0.56 and 0.79, respectively. But the first two of them have stellar mass ($10^{9.06}$ M$_\odot$ and $10^{9.02}$ M$_\odot$) right around the stellar mass boundary we set.
Our precision and recall for the local dwarf galaxies in the SAGA data is $95.95\%$ (71/74) and $67.62\%$ (71/105) respectively.  
Nevertheless, the classification model performs well and is consistent with expectations.



\subsection{Verification using ELVES}

The ELVES survey is designed to detect and characterize low-mass, dwarf satellite galaxies ($M_V < -9$ mag, $M_* \gtrsim 5\times 10^5$ M$_\odot$) around nearby, massive Milky-Way like hosts ($M_{K_s} < -22.1$ mag) in the Local Volume (LV; D $<$ 12 Mpc).
In total, they confirm 251 dwarf satellites \citep{Carlsten2020, Carlsten2021, ELVES2022} of 31 hosts with distances estimated using either measurements of surface brightness fluctuations (SBF), the tip of the red giant branch (TRGB) distances, or a spectroscopic redshift. 

The majority (27/31) of the hosts in ELVES survey are in the LS footprint, therefore 201 dwarf satellites are in the LS footprint. Again, these galaxies are not part of our training sample and so provide an independent test. Among the 135 satellites of those with a full feature set, 116 are dwarf satellites according to our definition (Table \ref{tab:training}). 
Applying our model, we classify 63 of the 135 as dwarf satellites. Of those 63 positive predictions, 61 are correctly identified. The two misclassifications have stellar mass of $10^{9.12}$ M$_\odot$ ($z_{spec}$ = 0.003 and $p_{\rm signal}$ = 0.78) and $10^{5.95}$ M$_\odot$ ($z_{spec}$ = 0.001 and $p_{\rm signal}$ = 0.99), the second one of which is around the stellar mass and redshift boundary of the signal. 
The precision and recall for local dwarf galaxies in the ELVES data set are $96.83\%$ (61/63) and $52.59\%$ (61/116), respectively. As for the SAGA data sample, the results are also consistent with our claims of high precision and moderate recall.

\subsection{Mass-SFR and mass-size relation}

Thanks to the redshift measurements from the DESI-EDR spectra, the distances for part of our dwarf sample can be determined, which allows us to investigate the physical properties of those dwarf galaxies. For example, the physical half light radius ($r_{1/2}$, in unit of kpc) can be directly calculated using the distances of those DESI-EDR galaxies and the measured half light radii in angular units. The stellar masses and SFRs are extracted from the catalog of \cite{Siudek2024}. We separate the galaxies into two subsample:  blue galaxies with $g-r \leq$ 0.7 and  red galaxies with $g-r >$ 0.7 \citep{Fernandez2012}.

\begin{figure*}[ht]
\begin{center}
\includegraphics[width = \textwidth]{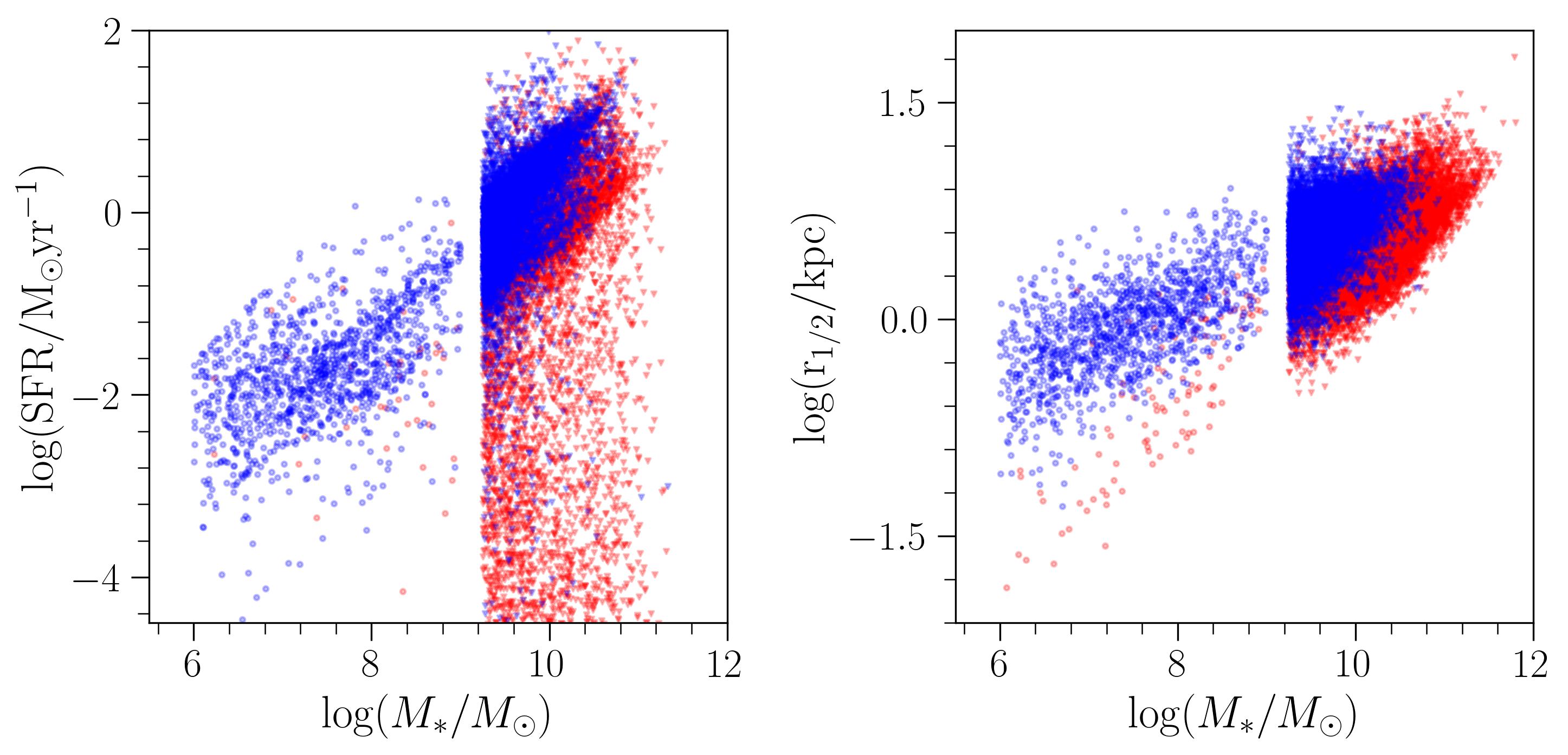}
\end{center}
\caption{$M_*$-SFR relation (left panel) and mass-size relation (right panel) for the dwarf galaxies as well as for the massive galaxies in the DESI-EDR database. The blue symbols represent the blue galaxies with $g-r <$ 0.7 and the red symbols stand for the red galaxies with $g-r >$ 0.7. The dots denote the dwarf galaxy sample and the triangles indicate the massive galaxies with stellar mass greater than $10^{9.2}$ M$_\odot$. }
\label{fig:stat}
\end{figure*}

In Figure \ref{fig:stat}, we present the stellar mass-SFR relation (left panel) and mass-size relation (right panel) for the dwarf galaxies, as well as for massive galaxies at $0.02 < z < 0.1$ with stellar mass greater than $10^{9.2}$ M$_\odot$, extracted from the DESI-EDR database. 
The majority dwarf galaxies in our sample are blue galaxies and follow the $M_*$-SFR relation of the massive galaxies \citep[e.g.,][]{Brinchmann2004, Peng2010}. It is also consistent with the predictions using EDGE (‘Engineering Dwarfs at Galaxy formation’s Edge’) simulations \citep{Rey2020} in the low mass range of $10^4 < M_*/M_\odot < 10^8$ \citep{Rey2025}. In general, the red dwarf galaxies are much more compact than the blue dwarf galaxies. 
The mass-size relation of the dwarf galaxy sample is similar to that of the massive galaxy sample \citep{Fernandez2013}.

\section{Dicussion} \label{sec:discussion}

We have demonstrated that the neural network classification model effectively separates the local dwarf galaxies from various contaminants. We confirm, using DESI-EDR public spectra, that $\sim$100\% of those classified as local dwarf galaxies are correctly identified. Even so, there is room for improvement in the recall (completeness).

\subsection{Discussion of the imbalance issue}
\label{sec:imbalance}

As we mentioned earlier, the numbers of sources in the various classes in the training sample are unequal (Table~\ref{tab:training}). The ratio of the number in the most populous class (higher redshift galaxies) to the least (local massive galaxies) is roughly $\sim 50:1$. Classification models trained on extremely imbalanced datasets tend to favor the majority class, making it easier for the sources in the minority classes to be incorrectly classified as belonging to the majority. Such errors lead to a lower recall for the minority class, which also indirectly affects the precision of the majority class.

There are multiple known ways to deal with imbalances in the training set. One is to reduce the entire training sample by randomly drawing a certain fraction to match the size of the minority class, which is called ``under-sampling". Another is to randomly duplicate objects or artificially generate new objects according to certain algorithms in the training sample such that the number of objects in the various classes matches that in the majority class, which is called ``over-sampling". 



\begin{table*}[ht]
\centering
\begin{tabular}{c|cc|ccc|ccc}
\hline
\multirow{2}{*}{\resizebox{3.5cm}{!}{\diagbox{\Huge{\bf Metrics}}{\Huge{\bf Methods}}}}
& \multicolumn{2}{c|}{NN class\_weight}                  
& \multicolumn{3}{c|}{Under-sampling} 
& \multicolumn{3}{c}{Over-sampling} 
\\
& Imbalance 
& \multicolumn{1}{c|}{Balance} 
& RUS        
& NM
& OSS       
& ROS      
& SMOTE     
& ADASYN     
\\ \hline

precision 
& 0.953         & 0.948           
& 0.869         & 0.893           & 0.954        
& 0.894         & 0.873           & 0.751         \\
recall    
& 0.761         & 0.708           
& 0.769         & 0.589           & 0.735        
& 0.851         & 0.869           & 0.878         \\
f1        
& 0.846         & 0.810           
& 0.816         & 0.710           & 0.830        
& 0.872         & 0.871           & 0.810         \\
AGF 
& 0.887         & 0.860           
& 0.884         & 0.790           & 0.875        
& 0.925         & 0.931           & 0.930         \\
\hline
\end{tabular}
\caption{\label{tab:balance} The evaluation metrics for the NN model on the imbalanced sample without resampling, using the {\it ``$class\_weight$"} parameter, and on the balanced sample for the test set.}
\end{table*}

We use the neural network classification algorithm implemented in the Keras framework to build a neural network model that incorporates a {\it $class\_weight$} parameter to adjust the weights applied to each class, with the aim of balancing the classes by giving a higher weight to the less common classes. The {\it $class\_weight$} algorithm assigns a weight to each sample based on the number of samples in each class, with the smaller the porportion of the class the greater the weight. The other commonly used under-sampling algorithms include: Random Under Sampler (RUS), Near Miss \citep[NM, ][]{Mani2003}, and One Sided Selection \citep[OSS, ][]{Kubat1997}. Those three algorithms adopt different methods to undersample the training sample. RUS is a fast and easy way to to randomly select samples from the majority class to balance the number of instances with those of the minority class. NM uses the k-nearest neighbors algorithm to remove majority class samples near the minority class. OSS rejects the majority class samples around the minority class and near the decision boundary, and those that have minimal impact on the model. 

The commonly used over-sampling algorithms include: Random Over Sampler (ROS), Synthetic Minority Oversampling Technique \citep[SMOTE,][]{Chawla2011}, and Adaptive Synthetic Sampling \citep[ADASYN,][]{He2008}. The approaches to over sample the dataset are different for these algorithms. ROS randomly samples and duplicates the minority class with replacement. SMOTE synthesizes new samples between the nearest neighbors of minority class samples in the feature space. And ADASYN focuses on synthesizing new samples between the nearest neighbors of minority class samples misclassified by the KNN classifier. 

In Table~\ref{tab:balance} we present the performance metrics based on the test set for the NN classification model without resampling, for the NN model using the {\it ``$class\_weight$"} parameter, and for the NN model with balanced samples using the above algorithms. For the $class\_weight$ balanced method, the $class\_weight$ parameter improves recall but reduces precision compared to the results from the imbalanced sample. Among the under-sampling balanced methods, the RUS method decreases the precision significantly, and the recall of the NM method also shows a significant decrement. The OSS performs as well as the imbalanced sample. This decline is reasonable because the under-sampling methods reduces the number of the majority samples significantly, leading to the loss of important information about the contaminants and shrinking the coverage of the feature spaces. Despite of such a small training set, we are able to obtain models with precision around 90\% using balanced dataset. This also reflects the fact that the local low-mass galaxies are separable from the contaminants in the high-dimensional feature spaces. Among the over-sampling balanced methods, the metrics obtained by the ROS and SMOTE we examine are quite consistent, while the precision of the ADASYN is significantly reduced. Compared to the metrics from the imbalanced training sample, they show increases in recall and decreases in precision. Although the performance of the over-sampling balanced methods is quite good in terms of trade-off between the precision and recall, one issue with these methods is that the over-sampled instances might not be representative and might be biased in the feature space. 

In conclusion, the methods explored to address the imbalanced training set act mostly to enhance the recall of the signal while sacrificing the precision. Our emphasis has been on precision 
and so, we do not implement any of these methods in producing our catalog of local dwarf galaxies.

\subsection{Improving the Machine Learning performance}
\label{sec:improvement}

The performance of any classification depends on the quality of the input data.
The typical uncertainty in stellar mass estimates reaches $\sim \pm 0.25$ dex, as shown in \cite{Siudek2024}. Although we have tried our best to separate signal from contaminants using the stellar mass estimated from SED fitting, there will be some mixing of signal and contaminants in the training sample due to stellar mass errors. The mixing within the different classes of contaminants is even more severe due to the scatter in stellar mass estimations. 
This problem will be mitigated using the future imaging surveys with deeper images and more photometric measurements, but will never be entirely resolved.

According to the galaxy number density predicted by UniverseMachine \citep{Behroozi2019}, there are millions of low-mass galaxies in the local Universe. However, our training sample (Table \ref{tab:training}) consists of a tiny fraction of those and therefore could easily not be representative. 
Large spectroscopic surveys such as DESI \citep{DESI1, DESI2} and Subaru Prime Focus Spectrograph \citep[PFS,][]{PFS2022} will expand the training sample significantly. Moreover, additional photometric measurements from future imaging surveys, particularly in different filters, will provide more useful features for the machine learning models.

Our training sample consists of all extended sources, with mean (median) half light radii of 4.57 (3.53) arcseconds. The features we are using in the machine learning model are the photometric measurements. 
Using the images themselves as the input to the machine learning algorithm (for example, a Convolutional Neural Network) might be even more effective than the model we develop in this study. In a recent study,  \cite{Tanoglidis2021a} implemented convolutional neural networks (CNNs) to search for low surface brightness galaxies using images from both DES and the Hyper Suprime-Cam Subaru Strategic Program \citep[HSC SSP,][]{HSC-SSP1, HSC-SSP2, HSC-SSP3}. They demonstrated that the evaluation scores of the CNN are significantly higher than those of the support vector machine (SVM) and RF algorithms, which use photometric measurements as features. 


\subsection{Future Imaging Survey}

There are multiple complementary imaging surveys on the way, with wavelength coverage from near-ultraviolet (NUV) to near-infrared (NIR), which will be the frontier in searches for local low-mass galaxies. There will also be more confirmed galaxies with across a wide stellar mass range at all redshifts in the near future from DESI \citep{DESI1, DESI2}, and PFS \citep{PFS2022}, expanding the size of training samples significantly. 

For example, the Chinese Space Station Telescope (CSST) with 2-meter aperture will cover $\sim$ 17000 square degree and have wavelength range of $2600-10000$ \AA \ with broad-band filters of NUV, {\it u, g, r, i, z, y}. The Roman Space Telescope with a 2.4-meter primary mirror \citep[RST,][]{RST2015} has wide wavelength coverage of $0.48-2.3$ microns with 8 broad-band filters. The near-infrared images will have much finer resolution and greater depth than the WISE {\it W1, W2} images, which will significantly enhance the machine learning performance. The Euclid Space Telescope \citep[EST,][]{Euclid2022, Euclid2024} is a 1.2-m-diameter telescope and will deliver data over 15,000 square degrees across a wavelength range of $0.55–2.02$ microns. The Early Release Observations (ERO) \citep{Cuillandre2024} of EST are astonishing. The Large Synoptic Survey Telescope \citep[LSST,][]{LSST} with an 8.4-meter primary mirror will deliver the best images in 6 filters for over 18,000 square degrees sky coverage, and the 5$\sigma$ point-source depth of the coadded maps will reach 27.5 for $r$-band \citep{LSST2019}. All those wide-field imaging surveys, especially the inclusion of the NUV and NIR photometric measurements, will significantly improve the performance of local dwarf galaxies searches. 

\section{Summary} \label{sec:sum}

We develop a machine learning model to identify local dwarf galaxies using only photometric measurements and present the result catalog of candidates. To do this,  we first obtain a representative training sample consisting of all spectrally confirmed signals and various categories of contaminants in the DESI Imaging Legacy Survey DR9 footprint. We use photometric measurements from Legacy Survey DR9 and WISE survey and constructed various features to train a machine learning model. In addition to the photometric measurements in the {\it g, r, z, W1}, and {\it W2} bands, we also implement the fluxes at different aperture radii (apflux) for those bands. And we find that the apfluxes, which encode the information of surface brightness profile, are critical for a successful machine learning model.

We compare the performance of several commonly used machine learning classification algorithms, including KNN, random forest, XGBoost, neural network and TabNet. The neural network classification model emerged as the optimal one. The neural network model relies more on the apfluxes while the random forest or XGBoost rely more on the colors and magnitudes. The performance of our neural network classification model can reach a high precision of 95.33\%, and a recall of 76.07\% in classifying local dwarf galaxies in our test set.
We discuss the issue that we face with an imbalanced training sample and find that different approaches to mitigating the imbalance do not result in significant improvements in our key metric.

We obtain a total of 39,819,821 sources after applying some basic selection criteria to the entire Legacy Survey DR9 catalog. We then apply our neural network classification model to these 39 million sources, we finally identify 112,859 as local low-mass dwarf galaxy candidates. Furthermore, our classification model is able to identify both star-forming and quiescent dwarf galaxies. Using DESI-EDR spectra, we confirm that 640 out of 1,255 spectroscopically confirmed local dwarf galaxies are successfully identified, corresponding to a comparably low recall (51.00\%) but an excellent precision of 100\%. We also test the model against the SAGA and ELVES galaxy samples, which are not in our training sample.
There we also find excellent precision, finding that $95.95\%$ and $96.83\%$, respectively, of the objects that we identify as dwarf galaxies are correctly identified. The recall can also reach 68\% and 53\%, respectively.



Our current model has room for improvement. The scatter of the stellar mass estimates, the comparably small size of the training sample, and the limited photometric measurements are all drawbacks of our current classification model. The future DESI survey will discover tens of thousands of new local low-mass galaxies, which will significantly expand the training sample size. 
New imaging surveys that introduce new measurements in the NUV or NIR will enhance the model performance significantly. The Legacy Survey DR10 currently includes the {\it i} band for $\sim$ half sky coverage of DR9, and the future data release including the {\it y} band is on the way as well. Future imaging surveys, such as CSST, RST, EST and LSST, will deliver much deeper, higher resolution images with many more photometric measurements (e.g., {\it NUV, y, H} and {\it K}) and will be the frontier for dwarf galaxies searching at all redshifts. A full exploitation of those surveys will require machine learning algorithms. 

\section{Acknowledgments}

HZ, GY, RW acknowledge financial support from the start-up funding of the Huazhong University of Science and Technology, the National Science Foundation of China grant (No. 12303007), and the CSST project on blue and extreme galaxies. DZ acknowledges financial support from NSF grant AST-2006785.  This research was supported by the Munich Institute for Astro-, Particle and BioPhysics (MIAPbP) which is funded by the Deutsche Forschungsgemeinschaft (DFG, German Research Foundation) under Germany's Excellence Strategy EXC-2094 390783311.

The Legacy Surveys consist of three individual and complementary projects: the Dark Energy Camera Legacy Survey (DECaLS; Proposal ID \#2014B-0404; PIs: David Schlegel and Arjun Dey), the Beijing-Arizona Sky Survey (BASS; NOAO Prop. ID \#2015A-0801; PIs: Zhou Xu and Xiaohui Fan), and the Mayall z-band Legacy Survey (MzLS; Prop. ID \#2016A-0453; PI: Arjun Dey). DECaLS, BASS and MzLS together include data obtained, respectively, at the Blanco telescope, Cerro Tololo Inter-American Observatory, NSF's NOIRLab; the Bok telescope, Steward Observatory, University of Arizona; and the Mayall telescope, Kitt Peak National Observatory, NOIRLab. Pipeline processing and analyses of the data are supported by NOIRLab and the Lawrence Berkeley National Laboratory (LBNL). The Legacy Surveys project is honored to be permitted to conduct astronomical research on Iolkam Du'ag (Kitt Peak), a mountain with particular significance to the Tohono O'odham Nation.

NOIRLab is operated by the Association of Universities for Research in Astronomy (AURA) under a cooperative agreement with the National Science Foundation. LBNL is managed by the Regents of the University of California under contract to the U.S. Department of Energy.

This project used data obtained with the Dark Energy Camera (DECam), which was constructed by the Dark Energy Survey (DES) collaboration. Funding for the DES Projects has been provided by the U.S. Department of Energy, the U.S. National Science Foundation, the Ministry of Science and Education of Spain, the Science and Technology Facilities Council of the United Kingdom, the Higher Education Funding Council for England, the National Center for Supercomputing Applications at the University of Illinois at Urbana-Champaign, the Kavli Institute of Cosmological Physics at the University of Chicago, Center for Cosmology and Astro-Particle Physics at the Ohio State University, the Mitchell Institute for Fundamental Physics and Astronomy at Texas A\&M University, Financiadora de Estudos e Projetos, Fundacao Carlos Chagas Filho de Amparo, Financiadora de Estudos e Projetos, Fundacao Carlos Chagas Filho de Amparo a Pesquisa do Estado do Rio de Janeiro, Conselho Nacional de Desenvolvimento Cientifico e Tecnologico and the Ministerio da Ciencia, Tecnologia e Inovacao, the Deutsche Forschungsgemeinschaft and the Collaborating Institutions in the Dark Energy Survey. The Collaborating Institutions are Argonne National Laboratory, the University of California at Santa Cruz, the University of Cambridge, Centro de Investigaciones Energeticas, Medioambientales y Tecnologicas-Madrid, the University of Chicago, University College London, the DES-Brazil Consortium, the University of Edinburgh, the Eidgenossische Technische Hochschule (ETH) Zurich, Fermi National Accelerator Laboratory, the University of Illinois at Urbana-Champaign, the Institut de Ciencies de l'Espai (IEEC/CSIC), the Institut de Fisica d’Altes Energies, Lawrence Berkeley National Laboratory, the Ludwig Maximilians Universitat Munchen and the associated Excellence Cluster Universe, the University of Michigan, NSF’s NOIRLab, the University of Nottingham, the Ohio State University, the University of Pennsylvania, the University of Portsmouth, SLAC National Accelerator Laboratory, Stanford University, the University of Sussex, and Texas A\&M University.

BASS is a key project of the Telescope Access Program (TAP), which has been funded by the National Astronomical Observatories of China, the Chinese Academy of Sciences (the Strategic Priority Research Program “The Emergence of Cosmological Structures” Grant \#XDB09000000), and the Special Fund for Astronomy from the Ministry of Finance. The BASS is also supported by the External Cooperation Program of Chinese Academy of Sciences (Grant \#4A11KYSB20160057), and Chinese National Natural Science Foundation (Grant \#12120101003, \#11433005).

The Legacy Survey team makes use of data products from the Near-Earth Object Wide-field Infrared Survey Explorer (NEOWISE), which is a project of the Jet Propulsion Laboratory/California Institute of Technology. NEOWISE is funded by the National Aeronautics and Space Administration.

The Legacy Surveys imaging of the DESI footprint is supported by the Director, Office of Science, Office of High Energy Physics of the U.S. Department of Energy under Contract No. DE-AC02-05CH1123, by the National Energy Research Scientific Computing Center, a DOE Office of Science User Facility under the same contract; and by the U.S. National Science Foundation, Division of Astronomical Sciences under Contract No. AST-0950945 to NOAO.

This research used data obtained with the Dark Energy Spectroscopic Instrument (DESI). DESI construction and operations is managed by the Lawrence Berkeley National Laboratory. This material is based upon work supported by the U.S. Department of Energy, Office of Science, Office of High-Energy Physics, under Contract No. DE–AC02–05CH11231, and by the National Energy Research Scientific Computing Center, a DOE Office of Science User Facility under the same contract. Additional support for DESI was provided by the U.S. National Science Foundation (NSF), Division of Astronomical Sciences under Contract No. AST-0950945 to the NSF’s National Optical-Infrared Astronomy Research Laboratory; the Science and Technology Facilities Council of the United Kingdom; the Gordon and Betty Moore Foundation; the Heising-Simons Foundation; the French Alternative Energies and Atomic Energy Commission (CEA); the National Council of Science and Technology of Mexico (CONACYT); the Ministry of Science and Innovation of Spain (MICINN), and by the DESI Member Institutions: www.desi.lbl.gov/collaborating-institutions. The DESI collaboration is honored to be permitted to conduct scientific research on Iolkam Du’ag (Kitt Peak), a mountain with particular significance to the Tohono O’odham Nation. Any opinions, findings, and conclusions or recommendations expressed in this material are those of the author(s) and do not necessarily reflect the views of the U.S. National Science Foundation, the U.S. Department of Energy, or any of the listed funding agencies.

We acknowledge the use of several open-source software packages. Specifically, we are grateful to the developers  of SciPy \citep{SciPy2020}, NumPy \citep{numpy2020}, AstroPy \citep{astropy}, Matplotlib \citep{Hunter2007}, pandas \citep{pandas}, scikit-learn \citep{scikit-learn}, and Keras \citep{chollet2015keras}. These tools provide critical support for data analysis, numerical computation, and machine learning applications throughout this project.

\bibliography{draft}{}
\bibliographystyle{aasjournal}



\end{document}